\documentstyle[12pt]{article}
\begin{document}
%
%
\newcommand{\fz}[1]{\mbox{$\left| \cal #1 \it (j,a)\right>$}}
\newcommand{\bfz}{\mbox{$\left| \rho (b,j,a) \right>$}}
\newcommand{\Pa}{{\rm I}\!{\rm P}}
\newcommand{\R}{{\rm I}\!{\rm R}}
\newcommand{\C}{{\rm l}\!\!\!{\rm C}}
\newcommand{\N}{{\rm I}\!{\rm N}}
\newcommand{\1}{{\rm 1}\hspace{-0.2mm}\!{\rm I}}
\def\cA{{\cal A}}
\def\cH{{\cal H}}
\def\cG{{\cal G}}
\def\cZ{{\cal Z}}
\def\cL{{\cal L}}
\def\cB{{\cal B}}
\def\cF{{\cal F}}
\def\cN{{\cal N}}
\def\bfz{{\bf z}}
\def\bfr{{\bf r}}
\def\bfu{{\bf u}}
\def\bfy{{\bf y}}
\def\bfj{{\bf j}}
\def\bfk{{\bf k}}
\def\bfA{{\bf A}}
\def\bfB{{\bf B}}
\def\bfE{{\bf E}}
\def\bfV{{\bf V}}
\def\R{{\rm I\kern-1.6pt{\rm R}}}
\def\al{\alpha}
\def\ga{\gamma}
\def\de{\delta}
\def\ka{\kappa}
\def\vphi{\varphi}
\def\p{\partial}
\def\no{\noindent}
\def\la{\lambda}
\def\La{\Lambda}
\def\rg{\rangle}
\def\lag{\langle}
\def\btd{\bigtriangledown}
\def\1{{\rm 1\kern-3pt{\rm I}}}
%
   \newcommand{\Bfz}{\mbox{$\left| \rho (b,j,a) \right>$}}
   \newcommand{\Bfa}[1]{\frac{\delta}{\delta b_{#1}}}
   \newcommand{\Bfaf}[2]{\partial^F_{#1}(#2)}
   \newcommand{\Bfaa}[2]{\partial^A_{#1}(#2)}
\def\kasten#1{\mathop{\mkern0.5\thinmuskip
                      \vbox{\hrule
                            \hbox{\vrule
                                  \hskip#1
                                  \vrule height#1 width 0pt
                                  \vrule}%
                            \hrule}%
                      \mkern0.5\thinmuskip}}
\def\yd#1{\mathop{
\setlength{\unitlength}{#1}
\begin{picture}(2.1,2.1)
\put(0,0){\line(1,0){1}}
\put(0,0){\line(0,1){2}}
\put(0,1){\line(1,0){2}}
\put(0,2){\line(1,0){2}}
\put(1,0){\line(0,1){2}}
\put(2,1){\line(0,1){1}}
\end{picture}}}%
\def\yta#1{\mathop{
\setlength{\unitlength}{#1}
\begin{picture}(2.1,2.1)
\put(0,0){\line(1,0){1}}
\put(0,0){\line(0,1){2}}
\put(0,1){\line(1,0){2}}
\put(0,2){\line(1,0){2}}
\put(1,0){\line(0,1){2}}
\put(2,1){\line(0,1){1}}
\put(0.4,1.28){\tiny 1}
\put(1.4,1.28){\tiny 2}
\put(0.4,0.28){\tiny 3}
\end{picture}}}%
\def\ytb#1{\mathop{
\setlength{\unitlength}{#1}
\begin{picture}(2.1,2.1)
\put(0,0){\line(1,0){1}}
\put(0,0){\line(0,1){2}}
\put(0,1){\line(1,0){2}}
\put(0,2){\line(1,0){2}}
\put(1,0){\line(0,1){2}}
\put(2,1){\line(0,1){1}}
\put(0.4,1.28){\tiny 1}
\put(1.4,1.28){\tiny 3}
\put(0.4,0.28){\tiny 2}
\end{picture}}}%
\def\ytc#1{\mathop{
\setlength{\unitlength}{#1}
\begin{picture}(2.1,2.1)
\put(0,0){\line(1,0){1}}
\put(0,0){\line(0,1){2}}
\put(0,1){\line(1,0){2}}
\put(0,2){\line(1,0){2}}
\put(1,0){\line(0,1){2}}
\put(2,1){\line(0,1){1}}
\put(0.4,1.28){\small 3}
\put(1.4,1.28){\small 1}
\put(0.4,0.28){\small 1}
\end{picture}}}%
\def\ytn#1{\mathop{
\setlength{\unitlength}{#1}
\begin{picture}(2.1,2.1)
\put(0,0){\line(1,0){1}}
\put(0,0){\line(0,1){2}}
\put(0,1){\line(1,0){2}}
\put(0,2){\line(1,0){2}}
\put(1,0){\line(0,1){2}}
\put(2,1){\line(0,1){1}}
\put(0.4,1.32){\small n}
\put(1.1,1.32){\small n+1}
\put(0.25,0.32){\small n-1}
\end{picture}}}
%
%
 \title{Mixed Symmetry Solutions
 of Generalized Three-Particle Bargmann-Wigner Equations in the
 Strong-Coupling Limit}
 \author{Winfried Pfister}
 \date{Institut f\"ur Theoretische Physik \\ Universit\"at T\"ubingen}
 \maketitle
%
%
\begin{abstract}
\begin{sloppypar}
 Starting from a nonlinear isospinor-spinor field equation,
 generalized three-particle Bargmann-Wigner equations are
 derived. In the strong-coupling limit, a special class of spin
 1/2 bound-states are calculated.
 These solutions which are antisymmetric with respect to all
 indices, have mixed symmetries in isospin-superspin space and in spin orbit
 space. As a consequence of this mixed symmetry, we get three solution
 manifolds. In appendix \ref{b}, table 2, these solution
 manifolds are interpreted as the three generations of leptons and quarks.
 This interpretation will be justified in a forthcoming paper.
\end{sloppypar}
\end{abstract}
%
%
\setlength{\abovedisplayskip}{0.35cm}
\setlength{\belowdisplayskip}{\abovedisplayskip}
\setlength{\abovedisplayshortskip}{0.35cm}
\setlength{\belowdisplayshortskip}{\abovedisplayshortskip}
\newtheorem{Satz}{Lemma}
\indent
 \section{Introduction} \label{sec1}
In various field theoretic models, three-fermion bound states are
assumed to play an important role. In general, ordinary
Schr\"odinger equations or Bethe Salpeter equations are used for
their calculation. As far as these models are based on nonlinear
spinor equations, for instance Nambu Lasinio Models or Heisenberg
Models, it is reasonable to apply generalized Bargmann Wigner
equations for the calculation of many-fermion bound states.
In previous papers we calculated two-fermion composites and three-fermion
composites by means of generalized B.~W.~equations \cite{17},\cite{prs},
\cite{scl}.
In \cite{16},\cite{17}, \cite{lqb} we showed that the effective dynamics
of these bound states leads to an unbroken SU(2)$\times$U(1) gauge theory
and that the three-fermion bound states may be interpreted as quarks and
leptons. However, because we used three-particle states with
symmetric
isospin-superspin dependence, we obtained isospin quartets instead of
isospin doublets which are required by phenomenology.
Furthermore, the three-particle solutions with symmetric
isospin-superspin dependence can only describe one generation of leptons and
quarks i.~e.~the three families have not been included up to now. \\
This paper is a first step in order to remove these drawbacks.
Starting from a nonlinear spinor-isospinor field equation we will
derive generalized Bargmann-Wigner equations. Then we will calculate
those solutions which have mixed symmetry in isospin-superspin
space and spin orbit space.
The calculations are performed in the strong-coupling limit. A physical
motivation for the application of the strong-coupling limit is given in
\cite{scl}, where also literature concerning the strong-coupling limit is
cited. In the mixed symmetric sector we not only get isospin doublets but
also three solution manifolds which can in principle describe the three
generations of leptons and quarks. The proof that the effective dynamics
of mixed symmetric three-fermion states and two-fermion composites includes
the three generations of leptons and quarks is postponed to a
forthcoming paper. \\
It should be mentioned that we can choose between the solution
manifold with symmetric isospin-superspin dependence and the
solution manifold with mixed symmetry in isospin-superspin space
by fine-tuning the coupling constant.
The value of the coupling constant can be chosen such that one of the two
solution manifolds acquire low masses whereas the masses of the
remaining solution manifold get high values i.~e.~it become unobservable. \\
The paper is organized as follows. In section \ref{sec2} we introduce the
subfermion model. In section \ref{sec3}, the three-particle equations are
given and the strong-coupling limit is performed. In section \ref{sec4} the
solutions of these three-particle equations are discussed for the case of
mixed symmetry in isospin-superspin space. In section \ref{sec5} a summary
is given. Finally we mention that some definitions and notations
used in this paper are given in appendix \ref{a} and \ref{b}.
\section{The model} \label{sec2}
The basic fermions of our model are
described by Dirac spinors which satisfy the following
nonlinear spinor-isospinor equation:

 \begin{equation}\left(i\gamma^\mu
 \partial_\mu-m\right)_{\alpha \beta}\delta_{AB}\psi_{\beta
 B}(x)=gV_{\alpha \beta \gamma \delta}^{ABCD}\psi_{\beta
 B}(x)\bar{\psi}_{\gamma C}(x)\psi_{\delta D}(x) \label{fg}
 \end{equation}

with

 \begin{eqnarray}
 V_{\alpha \beta \gamma \delta}^{ABCD} & = & \frac{1}{2} \sum_{h=1}^2
 \left( v_{\alpha \beta}^{h}\delta_{AB}v_{\gamma \delta}^{h}\delta_{CD}-
 v_{\alpha
 \delta}^{h}\delta_{AD}v_{\gamma \beta}^{h}\delta_{CB} \right) \nonumber
 \end{eqnarray}
 \begin{displaymath}v_{\alpha
 \beta}^{1}:=\delta_{\alpha \beta}\, ,  \, v_{\alpha
 \beta}^{2}:=i\gamma_{\alpha \beta}^5 \quad .
 \end{displaymath}

\vspace{0.4cm}
As leptons and quarks are assumed to be constituted by three of these
fermions we call them in the following subfermions, in contrast to the
fermions of the standard model. \\
If we use the charge conjugated spinor
$\varphi_{\alpha A }^c$ instead of the adjoint spinor
$\bar{\varphi}_{\alpha A }$ and furthermore introduce the definitions

  \begin{eqnarray}\phi_{\alpha \kappa } & := & \left\{
  \psi_{\alpha 1 } \, , \, \psi_{\alpha 2 } \, , \, \psi_{\alpha
  1 }^c \, ,\, \psi_{\alpha 2 }^c \right\}\nonumber \\
  Z & := & (\alpha,\kappa ) \nonumber \\
  D^\mu_{Z_1 Z_2} & := & i \gamma^\mu_{\alpha_1 \alpha_2}\delta_{\kappa_1
  \kappa_2} \nonumber \\
  m_{Z_1 Z_2} & := & m\delta_{\alpha_1 \alpha_2} \delta_{\kappa_1 \kappa_2}
   \nonumber \\
  U^h_{Z_1 Z_2 Z_3 Z_4} & := &  g  v^h_{\alpha_1
  \alpha_2}\left( v^h\, C \right)_{\alpha_3 \alpha_4}\delta_{\kappa_1
  \kappa_2}\gamma^5_{\kappa_3 \kappa_4} \quad , \nonumber \\
   & & \nonumber
  \end{eqnarray}

we can combine (\ref{fg}) and its charge conjugated equation
into one equation

  \begin{equation}\left( D_{Z_1 Z_2}^\mu \partial_\mu
  - m_{Z_1 Z_2} \right)\phi_{Z_2}({\bf r},t)=\sum_h U^h_{Z_1 \{ Z_2
  Z_3 Z_4 \}_{as}} \phi_{Z_2}({\bf r},t) \phi_{Z_3}({\bf r},t)
  \phi_{Z_4}({\bf r},t) \label{ag}
  \end{equation}

The canonical equal time anticommutator then reads
\begin{eqnarray} \left\{
\phi_{Z_1}({\bf r}_1,t),\phi_{Z_2}({\bf r}_2,t) \right\}
& = &  A_{Z_1 Z_2}\delta ({\bf r}_1-{\bf r}_2) \nonumber \\
A_{Z_1 Z_2} & := &  \gamma^5_{\kappa_1
\kappa_2} (C \gamma^0)_{\alpha_1 \alpha_2}\quad . \nonumber
\end{eqnarray}
We characterize the quantum states $\left|a\right>$ of the model (\ref{ag})
by the set of normal ordered matrix-elements for equal times $t$
\begin{equation}\varphi_n\left( {\bf r}_1,Z_1,\ldots ,{\bf r}_n,Z_n
\right| \left. a \right) := \left< 0 \right. \left| {\cal N}
\left\{ \phi_{Z_1}({\bf r}_1,t)\ldots \phi_{Z_n}({\bf r}_n,t)
\right\} \right| \left. a \right> \quad . \label{me}
\end{equation}
Introducing furthermore the generating
functional states for the normal transformed matrix-elements
\fz{F} with anticommuting sources
$j_Z({\bf r})$ and their corresponding duals $\partial_Z({\bf r})$,
we get as a compact formulation of the field dynamics the
functional energy equation

 \begin{eqnarray} (E_a-E_0)\fz{F} & = & K_{I_1
 I_2}j_{I_1}\partial_{I_2}\fz{F} \nonumber \\ & &  + \sum_h W_{I_1 I_2 I_3
 I_4}^h \left. \bigg\{ j_{I_1}\partial_{I_4}\partial_{I_3}\partial_{I_2}
 -3F^a_{I_4 I}j_{I_1}j_I\partial_{I_3}\partial_{I_2} \right. \nonumber \\ & &
 +\left(
 3F^a_{I_4 I}F^a_{I_3 I'}+\frac{1}{4}A_{I_4 I}A_{I_3 I'}\right)
 j_{I_1}j_I j_{I'}\partial_{I_2} \label{eeg} \\ & &
 \left. -\left( F^a_{I_4 I}F^a_{I_3 I'}+ \frac{1}{4}A_{I_4 I}A_{I_3 I'} \right)
 F^a_{I_2 I''}j_{I_1}j_I j_{I'}j_{I''} \right\}\fz{F}\, . \nonumber
 \end{eqnarray}

In (\ref{eeg}) we used the definitions

 \begin{eqnarray} K_{I_1 I_2} & := & iD^0_{Z_1 Z}\left( \vec{D}_{Z
 Z_2}\cdot\nabla_{{\bf r}_1}-m_{Z Z_2}\right)\delta \left(
 {\bf r}_1-{\bf r}_2 \right) \nonumber \\
 W^h_{I_1 I_2 I_3 I_4} & := & iD^0_{Z_1 Z} U^h_{Z \{ Z_2
  Z_3 Z_4 \}_{as}}\delta \left(
 {\bf r}_1-{\bf r}_2 \right)\delta \left(
 {\bf r}_1-{\bf r}_3 \right) \delta \left(
 {\bf r}_1-{\bf r}_4 \right) \nonumber \\
 A_{I_1 I_2} & := & A_{Z_1 Z_2}\delta \left(
 {\bf r}_1-{\bf r}_2 \right) \nonumber \\
 \fz{F} & := & \sum_{n=0}^\infty \frac{i^n}{n!}\, \left< 0 \right.
 \left| N \left\{ \phi_{I_1}\ldots \phi_{I_n}
 \right\} \right| \left. a \right>\, j_{I_1} \ldots j_{I_n}
 \left|0 \right>_f
 \quad , \nonumber
 \end{eqnarray}
where the functional state $\left| 0\right>_f$ satisfies $\partial_I
\left| 0\right>_f ={}_f\left< 0\right| j_I=
0$ and $ F^a $ in (\ref{eeg}) is the antisymmetric equal time limit of the
fermion field propagator. Furthermore it should be emphasized, that at
this stage of calculation we have not restricted ourselves onto Fock-space.
 \section{Three-particle equations} \label{sec3}
If projected in coordinate space, (\ref{eeg}) yields an infinite
set of coupled differential equations for the infinite set of
matrix-elements of normal ordered products of field operators. In
order to obtain generalized Bargmann-Wigner equations from this
set, we consider only the ``diagonal part'' of (\ref{eeg}) which
is given by
 \begin{equation}
 \omega \fz{F}^d = j_{I_1}K_{I_1 I_2}\partial_{I_2} \fz{F}^d-
 3 \sum_{h=1}^2  j_{I_1} W^h_{I_1 I_2 I_3 I_4}F^a_{I_4
 I'}j_{I'}\partial_{I_3}\partial_{I_2} \fz{F}^d\, .  \label{dt}
 \end{equation}
The non-diagonal part of (\ref{eeg}) is assumed to mediate the
interactions of the eigenstates of (\ref{dt}) \cite{16}. The
corresponding theory of effective interactions is
not the topic of this paper. Rather we want to investigate the
solutions of (\ref{dt}), in particular three-particle solutions.
By projecting with $ (\frac{1}{i})^3{}_f\! \left< 0\right|
\partial_{V_3}\partial_{V_2} \partial_{V_1}$ from the left
we obtain from (\ref{dt}) the set of equations
\begin{eqnarray}
\omega \, \varphi_{V_1 V_2 V_3}&=& K_{V_1 I}\,\varphi_{I V_2 V_3} +
K_{V_2 I}\,\varphi_{V_1 I V_3} + K_{V_3 I}\,\varphi_{V_1 V_2
I}\nonumber \\
& & -3\sum_{p\in S(3)}(-)^p\sum_{h=1}^2 W^h_{V_{p(1)}I_1 I_2
I_3}F_{I_3 V_{p(2)}}^a\,\varphi_{I_1 I_2
V_{p(3)}} \label{d1.1}
\end{eqnarray}
for the calculation of the three-particle amplitude
$\varphi_{V_1 V_2 V_3}$ which has to be antisymmetric with
respect to all indices. For the connection of (\ref{d1.1}) with ordinary
BW-equations we refer to \cite{scl} and the literature cited therein. \\
For a first draft, we consider eqn.~(\ref{d1.1}) in the strong-coupling limit
which is characterized by
\begin{equation}
K_{I I'}\to -im D^0_{I I'}\quad ,\quad -im D^0_{I I'}= m \delta_{\kappa
\kappa '}\gamma^0_{\alpha \alpha '}\delta ({\bf r}-{\bf
r}')\quad . \label{d1.2}
\end{equation}
For a physical motivation with respect to the use of the strong-coupling
limit see \cite{scl}. With (\ref{d1.2}), eqn.~(\ref{d1.1}) becomes
\begin{eqnarray}
\lefteqn{\omega \, \varphi_{V_1 V_2 V_3} +im\left( D^0_{V_1
I}\,\varphi_{I V_2 V_3} +D^0_{V_2 I}\,\varphi_{V_1 I V_3}
+ D^0_{V_3 I}\,\varphi_{V_1 V_2 I}\right) } \nonumber \\
& & = -3 \sum_{p\in S(3)}(-)^p\sum_{h=1}^2 W^h_{V_{p(1)}I_1 I_2
I_3}F_{I_3 V_{p(2)}}^a\,\varphi_{I_1 I_2
V_{p(3)}}\quad . \label{d2.1}
\end{eqnarray}
For the evaluation of (\ref{d2.1}) we need the explicit
form of $F^a$. As a first approximation we take for $F^a$ the
antisymmetric equal time limit of the {\sl free} fermion field
propagator, which we assume to be regularized in the
sense that $F^a\big|_{{\bf r}_1 = {\bf r}_2}<\infty$ (For a systematic
treatment of nonperturbative regularization see \cite{zt}).
Therefore we have
\begin{eqnarray}
F^a\left( \begin{array}{c}{\bf r}\\ \begin{array}{cc}\alpha_1 &
\alpha_2 \\ \kappa_1 &\kappa_2 \end{array} \end{array}\right) &
= & (\gamma_5)_{\kappa_1 \kappa_2}\left[i\vec{\gamma}C\cdot
\nabla_{\bf r} + mC\right]_{\alpha_1 \alpha_2}\, {\rm
Reg}\,\frac{-1}{2(2\pi)^3} \int\, d{\bf p}\,\frac{e^{-i{\bf p
r}}}{\sqrt{m^2+{\bf p}^2}} \nonumber \\
&=:& (\gamma_5)_{\kappa_1 \kappa_2}
F\left( \begin{array}{c}{\bf r}\\ \begin{array}{cc}\alpha_1 &
\alpha_2 \end{array} \end{array}\right)\nonumber \\
& =:& (\gamma_5)_{\kappa_1 \kappa_2}\left( \gamma^k C\, h^k
({\bf r}) + C\, s({\bf r})\right)_{\alpha_1 \alpha_2} \label{d6.2}
\end{eqnarray}
where $ {\bf r}:={\bf r}_1 -{\bf r}_2\,\, ,\,\, h^k(-{\bf
r})=-h^k({\bf r})$. Furthermore we mention that $s({\bf r})$
is a scalar function
which means that it is a function of ${\bf r}^2:= {\bf r}\cdot {\bf r}$.
Using Fierz identities, the vertex $\sum_{h=1}^2 W^h_{I I_1 I_2 I_3 }$
can be written as
\begin{eqnarray}
\lefteqn{ 3\sum_{h=1}^2 W^h_{I I_1 I_2 I_3}=g\,\delta ({\bf r}-{\bf r}_1)\,
\delta ({\bf r}-{\bf r}_2)\, \delta ({\bf r}-{\bf r}_3) \,\cdot }
 \nonumber \\
 & &
 \left\{ 4(\gamma^0\gamma^\mu )_{\alpha
\alpha_3}\frac{1}{4}(\gamma^\mu C)^\dagger_{\alpha_1
\alpha_2}(A^n\gamma_5)_{\kappa \kappa_3}\hat{A}^n_{\kappa_1
\kappa_2} + 4 (\gamma^0\gamma^\mu\gamma_5)_{\alpha
\alpha_3}\frac{1}{4}(\gamma^\mu\gamma_5C)^\dagger_{\alpha_2
\alpha_1}(S^m\gamma_5)_{\kappa \kappa_3}\hat{S}^m_{\kappa_1
\kappa_2}\right. \nonumber \\
& & \left. -4\gamma^0_{\alpha
\alpha_3}\frac{1}{4}C^\dagger_{\alpha_2 \alpha_1}(\gamma_5\cdot
\gamma_5)_{\kappa \kappa_3} (\gamma_5 )_{\kappa_1 \kappa_2}+4(\gamma^0\gamma_5
)_{\alpha \alpha_3}\frac{1}{4}(\gamma_5 C)^\dagger_{\alpha_2
\alpha_1}(\gamma_5\cdot\gamma_5 )_{\kappa \kappa_3}(\gamma_5)_{\kappa_1
\kappa_2}\right\} \label{d2.2}
\end{eqnarray}
where $A^n$ or $S^m$ is an arbitrary complete set of antisymmetric or
symmetric $4\times 4$ matrices, and
$\hat{A}^n$ or $\hat{S}^m$ resp.~are the corresponding duals which
satisfy the following completeness and orthogonality relations:
\begin{displaymath}
-{\rm tr}\left[\hat{A}^n A^{n'}\right]=\delta_{n n'}\quad ,
\quad
A^n_{\kappa_1 \kappa_2} \hat{A}^n_{\kappa_1'
\kappa_2'}=\frac{1}{2}\left( \delta_{\kappa_1 \kappa_1'}\delta_{\kappa_2
\kappa_2'}- \delta_{\kappa_1 \kappa_2'}\delta_{\kappa_2
\kappa_1'}\right)
\end{displaymath}
\begin{displaymath}
{\rm tr}\left[\hat{S}^m S^{m'}\right]=\delta_{m m'}\quad , \quad
S^m_{\kappa_1 \kappa_2} \hat{S}^m_{\kappa_1'
\kappa_2'}=\frac{1}{2}\left( \delta_{\kappa_1 \kappa_1'}\delta_{\kappa_2
\kappa_2'}+ \delta_{\kappa_1 \kappa_2'}\delta_{\kappa_2 \kappa_1'}\right)
\quad .
\end{displaymath}
\section{Calculation of three-particle states in the mixed symmetric
sector} \label{sec4}
In this section, we discuss those spin 1/2 solutions of (\ref{d2.1}) which
possess mixed symmetry in isospin-superspin space. In addition, we restrict
ourselves to solutions $\varphi$ which satisfy the condition
\begin{equation}
(\gamma_5)_{\kappa_1\kappa_2}\,\varphi\left(\begin{array}{ccc}
\bfr_1&\bfr_2&
\bfr_3\\\alpha_1&\alpha_2&\alpha_3\\\kappa_1&\kappa_2&\kappa_3
\end{array}\right)=0 \label{I3.1}
\end{equation}
As will be shown below, the constraint
(\ref{I3.1}) enables us to completely separate the determination of the
isospin-superspin part from the calculation of the spin orbit part of the
wave function. This is in full analogy to the case of symmetric
isospin-superspin dependence (see \cite{scl}). \\
The ansatz for an antisymmetric function $\varphi_{I_1
I_2 I_3}$ with mixed symmetry in isospin-superspin space reads \cite{kjs}:
\begin{eqnarray}
\left| \varphi^{j,a} \right> &=& C_{11}\left| j \right> \otimes
C_{22}\left|\Phi^a \right>- C_{21}\left| j\right>\otimes
C_{12}\left|\Phi^a \right> \nonumber \\
&&+ C_{22}\left| j\right>\otimes C_{11}\left| \Phi^a \right>-
C_{12}\left| j\right>\otimes C_{21}\left|\Phi^a \right> \label{I3.2}
\end{eqnarray}
The Young-operators $C_{ik}$ are defined in appendix \ref{a}.
The quantum number $a$ represents the $J=1/2$ spin quantum numbers whereas
the quantum number $j$ of the isospin state $\left| j\right>$ combines
the isospin and fermion quantum numbers. The possibility to classify the
states according to isospin and fermion quantum numbers is a consequence
of the global SU(2)$\times$U(1)-invariance of equation (\ref{fg}) or
(\ref{d2.1}) respectively. Furthermore we have used the Dirac bracket
formulation. For instance
\jot1mm
\begin{eqnarray*}
\bigg\{ \left< \kappa_1\kappa_2\kappa_3\right|\otimes \left<
\begin{array}{ccc}\bfr_1&\bfr_2&\bfr_3\\
\alpha_1&\alpha_2&\alpha_3 \end{array}\right| \bigg\} \,
\left|\varphi \right>& =&\varphi\left(\begin{array}{ccc}
\bfr_1&\bfr_2&
\bfr_3\\\alpha_1&\alpha_2&\alpha_3\\\kappa_1&\kappa_2&\kappa_3
\end{array}\right) \\
\bigg\{ \left< \kappa_1\kappa_2\kappa_3\right|\otimes \left<
\begin{array}{ccc}\bfr_1&\bfr_2&\bfr_3\\
\alpha_1&\alpha_2&\alpha_3 \end{array}\right| \bigg\}\,
\bigg\{ \left| j\right>\otimes \left|\Phi \right>\bigg\} &=&
\Theta^j_{\kappa_1\kappa_2\kappa_3} \, \Phi  \left(\begin{array}{ccc}
\bfr_1&\bfr_2&
\bfr_3\\\alpha_1&\alpha_2&\alpha_3 \end{array}\right)
\end{eqnarray*}
\jot1mm
with
$\Theta^j_{\kappa_1\kappa_2\kappa_3}=\left<
\kappa_1\kappa_2\kappa_3\right. \left| j\right>$,
etc. \\
In general, we must take into account the possibility of degeneracy.
Therefore we have to replace in (\ref{I3.2}) the states $\left| j\right>$ or
$\left| \Phi^a\right>$ resp.~by the linear combinations
$a_s\,\left| j,s\right>$ or
$b_r\,\left| \Phi^a_r\right>$ respectively, where the degeneracy indices
$s,r$ enumerate the states belonging to the same quantum number.
With these replacements we get from (\ref{I3.2}):
\begin{eqnarray}
\left| \varphi \right> &=& a_s\, b_r\,\bigg\{ C_{11}\left| j,s \right> \otimes
C_{22}\left|\Phi^a_r \right>- C_{21}\left| j,s\right>\otimes
C_{12}\left|\Phi^a_r \right> \nonumber \\
&&+ C_{22}\left| j,s\right>\otimes C_{11}\left| \Phi^a_r \right>-
C_{12}\left| j,s\right>\otimes C_{21}\left|\Phi^a_r \right> \bigg\}
\label{ansatz}
\end{eqnarray}
In the mixed symmetric state space, the unit operator is given by the sum
of the projection operators $C_{11}$ and $C_{22}$. This yields a decomposition
of the mixed symmetric state space according to $H_{\rm mixed}=H_{11}\oplus
H_{22}$. If we require the states $C_{11}\left| j,s\right>$ to be complete,
i.~e.~every state of $H_{11}$ with quantum number $j$ can uniquely be written
as $\alpha_s C_{11}\left| j,s\right>$, then it can be proven that we get
a complete set of states in $H_{22}$ with the help of the step operator
$C_{21}$:
\begin{equation}
C_{22}\left| j,s\right> = \alpha_{s s'}\, C_{21}C_{11}\left| j,s'\right>
\stackrel{(\ref{p3.2})}{=}\alpha_{s s'}\, C_{21}\left| j,s'\right> \label{l1}
\end{equation}
With the analog requirement for the state $C_{22}\left| \Phi^a_r\right>$, we
have
\begin{equation}
C_{11}\left| \Phi^a_r \right> = \beta_{r r'}\, C_{12}C_{22}
\left| \Phi^a_{r'}\right>
=  \beta_{r r'}\, C_{12}\left| \Phi^a_{r'} \right>\label{l2}
\end{equation}
If we substitute (\ref{l1}) and (\ref{l2}) into (\ref{ansatz}), we get
\begin{eqnarray*}
\left|\varphi^{j,a}\right>&=& a_s b_r \bigg\{
C_{11}\left|j,s\right>\otimes C_{22}\left|\Phi^a_r\right>-
C_{21}\left|j,s\right> \otimes C_{12}\left|\Phi^a_r\right> \\
&&+\alpha_{ss'}\beta_{rr'}\,\,C_{21}\left|j,s'\right> \otimes
C_{12}\left|\Phi^a_{r'}\right>- \alpha_{ss'} \beta_{rr'}
\,\,C_{11}\left|j,s'\right>\otimes C_{22}\left|\Phi^a_{r'}\right> \bigg\}
\end{eqnarray*}
which can be written in the form
\jot1mm
\begin{eqnarray}
\left|\varphi^{j,a}\right>&=& \left( a_s b_r -a_{s'}\alpha_{s' s}
b_{r'}\beta_{r' r}\right)\,\,
C_{11}\left|j,s\right>\otimes C_{22}\left|\Phi^a_r\right> \nonumber \\
&&- \left( a_s b_r -a_{s'}\alpha_{s' s}
b_{r'}\beta_{r' r}\right) \,\,
C_{21}\left|j,s\right>\otimes C_{12}\left|\Phi^a_r\right>\nonumber \\
&=:&t_{sr}\, \Bigl( C_{11}\left|j,s\right>\otimes C_{22}\left|\Phi^a_r\right>
- C_{21}\left|j,s\right>\otimes C_{12}\left|\Phi^a_r\right>\Bigr) \label{I8.1}
\end{eqnarray}
\jot0mm
So far we have not achieved any simplification with respect to (\ref{ansatz}).
However, it is shown in appendix \ref{b} that due to the requirement
(\ref{I3.1}) there is no degeneracy in the isospin-superspin
space i.~e with (\ref{I3.1}) we have
\begin{eqnarray*}
C_{11}\left| j,s\right>&\to& C_{11}\left| j\right> \\
a_s\, ,\,\alpha_{s s'} &\to& a\, ,\, \alpha \\
\left( a\, b_r -a\alpha\, b_{r'}\beta_{r' r}\right)&\to&\eta_r
\end{eqnarray*}
Hence, we infer from (\ref{I8.1}):
\begin{equation}
\left|\varphi^{j,a}\right>=\eta_r\,\,
C_{11}\left|j\right>\otimes C_{22}\left|\Phi^a_r\right>
- \eta_r\,\, C_{21}\left|j\right>\otimes
C_{12}\left|\Phi^a_r\right>\label{I8.2}
\end{equation}
The ansatz (\ref{I8.2}) is indeed a simplification in comparison to the
ansatz (\ref{ansatz}) because in this ansatz we only have coefficients
$\eta_s$ with one index instead of coefficients $t_{rs}$ with two indices.
Before we use this ansatz to evaluate (\ref{d2.1}), we discuss a further
consequence of (\ref{I3.1}).\\
By projecting in (\ref{eeg}) with ${}_f\!\left< 0 \right|\partial_I$,
we see that a condition for
$\varphi_{I_1 I_2 I_3}$ to describe a genuine three-particle state and
not a polarization cloud of one particle, is given by
\begin{equation}
\sum_{h=1}^2 W^h_{I I_1 I_2 I_3}\varphi_{I_1 I_2 I_3} \stackrel{!}{=}0
\label{I9.1}
\end{equation}
In the following we demonstrate that the condition (\ref{I3.1}) is sufficient
for the fulfillment of (\ref{I9.1}).
After substitution of (\ref{I8.2}) into (\ref{I9.1}) we get:
\begin{eqnarray}
\eta_r\,(\gamma^0\gamma^{\mu})_{\alpha
\alpha_3}\frac{1}{4}(\gamma^{\mu}C)^{\dagger}_{\alpha_1\alpha_2}
(\gamma_5)_{\kappa_2\kappa_3}
(C_{11}\Theta^j)_{\kappa\kappa_2\kappa_3}
(C_{22}\Phi^a_r)_{\alpha_1\alpha_2\alpha_3} && \nonumber \\
- \eta_r\,(\gamma^0\gamma^{\mu}\gamma_5)_{\alpha
\alpha_3}\frac{1}{4}(\gamma^{\mu}\gamma_5 C)^{\dagger}_{\alpha_1\alpha_2}
(\gamma_5)_{\kappa_2\kappa_3}
(C_{21}\Theta^j)_{\kappa\kappa_2\kappa_3}
(C_{12}\Phi^a_r)_{\alpha_1\alpha_2\alpha_3} &\stackrel{!}{=}&0 \label{I10.1}
\end{eqnarray}
where we already have taken into account that due to (\ref{I3.1}), the last
two terms on the right hand side of (\ref{d2.2}) do not give any contribution
in (\ref{I9.1}). In order to further simplify (\ref{I10.1}), we have to
separate the isospin-superspin part from the spin part. Therefore we need
\begin{Satz}\label{Satz3}
\begin{displaymath}
(\gamma_5)_{\kappa_2\kappa_3}(C_{11}\Theta^j)_{\kappa_1\kappa_2\kappa_3}
=\sqrt{3} (\gamma_5)_{\kappa_2\kappa_3}
(C_{21}\Theta^j)_{\kappa_1\kappa_2\kappa_3}
\end{displaymath}
\end{Satz}
{\bf Proof:} \\
Due to $(\gamma_5)_{\kappa_1\kappa_2} =
(\gamma_5)_{\kappa_2\kappa_1}$ we have
\begin{displaymath}
(\gamma_5)_{\kappa_2\kappa_3}
(C_{11}\Theta^j)_{\kappa_1\kappa_2\kappa_3}
=(\gamma_5)_{\kappa_2\kappa_3}
(P_{23}C_{11}\Theta^j)_{\kappa_1\kappa_2\kappa_3} \quad ,
\end{displaymath}
where the relation
$P_{23}C_{11}\stackrel{(\ref{p3.3})}{=}1/2\, C_{11}
+1/2\,\sqrt{3}\,C_{21}$ completes the proof. \hfill $\Box$ \\
\\[-3mm]
With lemma \ref{Satz3}, condition (\ref{I10.1}) is simplified and yields
\begin{eqnarray}
(\gamma_5)_{\kappa_1\kappa_2}(C_{11}\Theta^j)_{\kappa\kappa_1\kappa_2}
\, \cdot \, \bigg\{
\eta_r\,(\gamma^0\gamma^{\mu})_{\alpha
\alpha_3}\frac{1}{4}(\gamma^{\mu}C)^{\dagger}_{\alpha_1\alpha_2}
(C_{22}\Phi^a_r)_{\alpha_1\alpha_2\alpha_3}&& \nonumber \\
- \eta_r\, \frac{1}{\sqrt{3}}\, (\gamma^0\gamma^{\mu}\gamma_5)_{\alpha
\alpha_3}\frac{1}{4}(\gamma^{\mu}\gamma_5 C)^{\dagger}_{\alpha_1\alpha_2}
(C_{12}\Phi^a_r)_{\alpha_1\alpha_2\alpha_3}
 \bigg\}& \stackrel{!}{=}& 0 \label{I11.1}
\end{eqnarray}
We prove that (\ref{I11.1}) is automatically fulfilled if (\ref{I3.1})
is postulated. Substituting the ansatz (\ref{I8.2}) into (\ref{I3.1}) yields
\begin{eqnarray}
(\gamma_5)_{\kappa_1 \kappa_2}
(C_{11}\Theta^j)_{\kappa_1\kappa_2\kappa_3}&=&0 \label{I11.2}\\
(\gamma_5)_{\kappa_1 \kappa_2}
(C_{21}\Theta^j)_{\kappa_1\kappa_2\kappa_3}&=&0 \label{I11.3}
\end{eqnarray}
Equation (\ref{I11.2}) is trivially fulfilled due to the
antisymmetry of $(C_{11}\Theta^j)$ in the first two indices
whereas the requirement (\ref{I11.3}) is a genuine restriction
to the $\Theta^j$. Therefore the requirement (\ref{I11.3}) is fully
equivalent to the requirement (\ref{I3.1}). In order to get the connection
between (\ref{I11.3}) and (\ref{I11.1}), we prove the following
\begin{Satz}\label{Satz4}
\begin{displaymath}
(\gamma_5)_{\kappa_2\kappa_3}(C_{21}\Theta^j)_{\kappa_2\kappa_3\kappa_1}
=-\frac{2}{\sqrt{3}} (\gamma_5)_{\kappa_2\kappa_3}
(C_{11}\Theta^j)_{\kappa_1\kappa_2\kappa_3}
\end{displaymath}
\end{Satz}
{\bf Proof:}\\

\begin{eqnarray*}
(\gamma_5)_{\kappa_2\kappa_3}(C_{11}\Theta^j)_{\kappa_1\kappa_2\kappa_3}&
=&
(\gamma_5)_{\kappa_2\kappa_3}(P_{13}P_{13}C_{11}\Theta^j)_{\kappa_1
\kappa_2 \kappa_3} \\
&\stackrel{(\ref{p3.4})}{=}& (\gamma_5)_{\kappa_2\kappa_3}\left[
P_{13}\left(
\frac{1}{2}C_{11}-\frac{\sqrt{3}}{2}C_{21}\right)
\Theta^j\right]_{\kappa_1\kappa_2\kappa_3} \\
&=& (\gamma_5)_{\kappa_2\kappa_3}\left[
\left( \frac{1}{2}C_{11}-\frac{\sqrt{3}}{2}C_{21}\right)
\Theta^j\right]_{\kappa_3\kappa_2\kappa_1} \\
&=&
-\frac{\sqrt{3}}{2}(\gamma_5)_{\kappa_2\kappa_3}
(C_{21}\Theta^j)_{\kappa_2\kappa_3\kappa_1}
\hspace*{4.9cm}\Box
\end{eqnarray*}
Due to lemma \ref{Satz4}, we see that (\ref{I11.1}) is a consequence
of the requirement (\ref{I3.1}) i.e.~those functions
which fulfill (\ref{I3.1}) do not describe polarization clouds. \\
Keeping in mind that due to (\ref{I3.1}) we can neglect the last two terms
on the right hand side of (\ref{d2.2}), we get after substituting
(\ref{I8.2}) into (\ref{d2.1}) an eigenvalue equation with the
following structure
\begin{eqnarray}
\left\{\omega-m\left(\gamma_1^0+\gamma_2^0+\gamma_3^0\right)\right\}\left|
\varphi^{j,a}\right> &=& \eta_r \, C^a\circ\bigg\{ C_{11}\left|
j\right>\otimes \hat{O_1}C_{22}\left| \Phi^a_r \right> \nonumber
\\
&& -C_{21}\left| j\right>\otimes\hat{O_2}C_{12}\left| \Phi^a_r
\right>\bigg\} \label{I13.1}
\end{eqnarray}
where $\hat{O}_{1,2}$ are well defined operators.
The operator $C^a$ is the usual antisymmetrisizer. It's
Kronecker decomposition is given by \cite{kjs},p37,prop.~2.~58:
\begin{eqnarray}
C^a&=&C^a\times C^s + C^s\times C^a +\frac{1}{2}\big( C_{11}\times
C_{22}-C_{21}\times C_{12}+C_{22}\times C_{11}\nonumber \\
&&-C_{12}\times C_{21}\big) \label{p4.1}
\end{eqnarray}
With the help of (\ref{p4.1}),(\ref{I14.2}) and (\ref{p3.2})
we get from (\ref{I13.1}):
\begin{eqnarray}
\lefteqn{
\left\{\omega-m\left(\gamma_1^0+\gamma_2^0+\gamma_3^0\right)\right\}\left|
\varphi^{j,a}\right>=
} \nonumber \\
&&\eta_r \frac{1}{2} \bigg\{  C_{11}\left| j\right>\otimes
C_{22}\hat{O_1}C_{22}\left| \Phi^a_r \right> -  C_{21}\left|
j\right>\otimes C_{12}\hat{O_1}C_{22}\left| \Phi^a_r \right>
\nonumber \\
&&-C_{21}\left| j\right>\otimes C_{11}\hat{O_2}C_{12}\left| \Phi^a_r
\right> + C_{11}\left| j\right>\otimes C_{21}\hat{O_2}C_{12}\left| \Phi^a_r
\right> \label{I14.3}
\end{eqnarray}
If we multiply in (\ref{I14.3}) from the left with
\begin{displaymath}
 \left<
j\right|( C_{11}\times \1 ) \quad {\rm
or}\quad  \left<
j\right| ( C_{12}\times \1 )\quad ,
\end{displaymath}
we get with the normalization of the states $\left| j\right>$
(see \ref{norm}):
\begin{equation}
\left\{\omega-m\left(\gamma_1^0+\gamma_2^0+\gamma_3^0\right)\right\}
\eta_r C_{22} \left| \Phi^a_r \right> = \frac{1}{2} \left\{
C_{22}\hat{O_1} + C_{21}\hat{O_2}C_{12}\right\} \eta_r
C_{22}\left|\Phi^a_r\right> \label{I15.1}
\end{equation}
In (\ref{I15.1}) we have no isospin-superspin dependence
i.~e.~we have completely separated the calculation of the
isospin-superspin part from the spin orbit part of the three-particle
states. If $\eta_r C_{22}\left| \Phi^a_r\right>$
is given, i.~e.~is calculated from (\ref{I15.1}), we get the complete state
$\left|\varphi^{j,a}\right>$ with the help of the relation
\begin{equation}
\left| \varphi^{j,a}\right>=\Big( \1\times\1 -C_{21}\times
C_{12}\Big) \Big\{ C_{11}\left| j\right>\otimes \eta_r
C_{22}\left| \Phi^a_r\right>\Big\}\label{I15.2}
\end{equation}
where the states $C_{11}\left| j\right>$ which fulfill
\begin{equation}
(\gamma_5)_{\kappa_2\kappa_3}\left<
\kappa_1\kappa_2\kappa_3\right|C_{11} \left| j\right>=
(\gamma_5)_{\kappa_2\kappa_3}\left( C_{11} \Theta^j
\right)_{\kappa_1\kappa_2\kappa_3} =0
\end{equation}
are given in appendix \ref{b}. \\
If projected in configuration space, equation (\ref{I15.1}) reads:
\jot2mm
\begin{eqnarray}
\lefteqn{
\bigg\{ \omega-m\left(
(\gamma^0 )_{\alpha_1 \beta_1} \delta_{\alpha_2 \beta_2}
\delta_{\alpha_3 \beta_3} +
(\gamma^0 )_{\alpha_2 \beta_2} \delta_{\alpha_1 \beta_1}
\delta_{\alpha_3 \beta_3} +
(\gamma^0 )_{\alpha_3 \beta_3} \delta_{\alpha_2 \beta_2}
\delta_{\alpha_1 \beta_1}
\right) \bigg\}\cdot
}\nonumber \\
&& \cdot  C_{22} (\bfr ,\beta )\circ
\eta_r \Phi_r  \left(\begin{array}{ccc}
\bfr_1&\bfr_2&
\bfr_3\\ \beta_1&\beta_2&\beta_3 \end{array}\right)
 = \frac{g}{2} \, \int d\bfz_1 \,
d\bfz_2\, d\bfz_3\, \bigg\{ C_{22}(\bfr,\alpha )\circ
\nonumber \\
&& (\gamma^0\gamma^\mu )_{\alpha_1 \beta} \, \,
F\left( \begin{array}{c}\bfr_1-\bfr_2\\ \begin{array}{cc}\beta &
\alpha_2 \end{array} \end{array}\right)
 (\gamma^\mu C )_{\beta_1 \beta_2}^\dagger \delta_{\alpha_3 \beta_3}
\, \delta (\bfr_1 -\bfz_1 ) \, \delta (\bfr_1 -\bfz_2 )
\, \delta (\bfr_1 -\bfz_3 ) \nonumber \\
&& + \, C_{21} (\bfr ,\alpha )\circ
(\gamma^0 \gamma^\mu \gamma_5 )_{\alpha_1 \beta} \, \,
F\left( \begin{array}{c}\bfr_1-\bfr_2\\ \begin{array}{cc}\beta &
\alpha_2 \end{array} \end{array}\right)(\gamma^\mu \gamma_5 C )_{\beta_2
\beta_1}^\dagger \delta_{\alpha_3 \beta_3} \cdot \nonumber \\
&& \delta (\bfr_1 -\bfz_1 ) \, \delta (\bfr_1 -\bfz_2 )
\, \delta (\bfr_1 -\bfz_3 )\, C_{12} (\bfz, \beta ) \circ \bigg\} \,
C_{22} (\bfz,\beta ) \circ \eta_r \Phi_r  \left(\begin{array}{ccc}
\bfz_1&\bfz_2&
\bfz_3\\\beta_1&\beta_2&\beta_3 \end{array}\right) \label{G7.1}
\end{eqnarray}
\jot0mm
The notation $C_{ik}(\bfr ,\alpha )$ indicates that the Young-operator
$C_{ik}$ act on the indices $\alpha,\bfr$. \\
Equation (\ref{G7.1}) can in principle be solved \cite{scl}.
However, because we are only interested
in the structure of the solution rather than in it's detailed form,
we only consider (\ref{I15.1}) for $\bfk =0\,\, ,\,\,
\bfr_1=\bfr_2=\bfr_3=\bfr$ (see also \cite{scl}). In this case we can make
the ansatz
\begin{equation}
\eta_r C_{22}\Phi^a_r\left(\begin{array}{ccc}
\bfr_1&\bfr_2&
\bfr_3\\\alpha_1&\alpha_2&\alpha_3
\end{array}\right)\bigg|_{\def\arraystretch{0.5}\scriptstyle
\begin{array}{l}\scriptstyle \bfk=0 \\ \tiny \bfr_1=\bfr_2
=\bfr_3\end{array} \def\arraystretch{1}}=\eta_r\, \Omega^a_r
(\alpha_1\alpha_2\alpha_3) \label{I17.1}
\end{equation}
Because we are interested in spin-1/2-solutions, we need a complete set of
multispinors of the third kind in $H_{22}$ which describe spin $1/2$-states.
Without proof we give the following lemma:
\begin{Satz}\label{Satz5}
The multispinors of the third kind which describe spin $1/2$
and which are eigenstates of $C_{22}$ with eigenvalue 1
(i.~e.~are elements of $H_{22}$), are
unique linear combinations of the following three linear
independent multispinors
\begin{eqnarray*}
\hat{\Omega}^a_1&=&\frac{K_{\mu}K^{\nu}}{K^2}\Big\{
\gamma^{\mu}C\otimes\gamma_{\nu}\cdot \hat{\chi}^a(K) + \Sigma^{\mu
\rho}C\otimes \Sigma_{\nu\rho}\cdot\hat{\chi}^a
(K)\Big\}\label{I18.1} \\
\hat{\Omega}^a_2&=&\gamma^{\mu}C\otimes\gamma_{\mu}\cdot\hat{\chi}^a(K)
\label{I18.2} \\
\hat{\Omega}_3^a&=&\Sigma^{\mu\nu}C\otimes\Sigma_{\mu\nu}\cdot\hat{\chi}^a(K)
\quad , \label{I18.3}
\end{eqnarray*}
where we have used the notation $(A\otimes\psi
)_{\alpha_1\alpha_2\alpha_3}:=A_{\alpha_1\alpha_2}
\,\psi_{\alpha_3}$. Furthermore the Diracspinor
$\hat{\chi}^a(K)$ is assumed to fulfill
$(\gamma^{\mu}K_{\mu}-\sqrt{K^2})\hat{\chi}^a =0$.
\end{Satz}
Because we are interested in the case $\bfk =0$, we have to perform the
limit $\bfk\to 0$ in lemma \ref{Satz5}:
\jot1mm
\begin{eqnarray}
\hat{\Omega}^a_1&\stackrel{\bfk =0}{\to}&\Omega^a_1=
\gamma^0C\otimes\gamma^0\cdot
\chi^a + \Sigma^{0k} \otimes \Sigma_{0k}\cdot\chi^a \\
\hat{\Omega}^a_2 &\stackrel{\bfk =0}{\to}& \Omega^a_2= \gamma^{\mu}C\otimes
\gamma_{\mu}\cdot\chi^a \\
\hat{\Omega}^a_3 &\stackrel{\bfk =0}{\to}& \Omega^a_3= \Sigma^{\mu\nu}C\otimes
\Sigma_{\mu\nu}\cdot \chi^a
\end{eqnarray}
\jot0mm
where
\begin{displaymath}
\hat{\chi}^a\stackrel{\bfk =0}{\to}\chi^a\in\left\{
\left( \begin{array}{c} 1\\0\\0\\0\end{array}\right)\, ,\, \left(
\begin{array}{c} 0\\1\\0\\0\end{array}\right) \, ,\, \left(
\begin{array}{c} 0\\0\\1\\0\end{array}\right) \, ,\, \left(
\begin{array}{c} 0\\0\\0\\1\end{array}\right) \right\}
\end{displaymath}
To solve (\ref{I15.1}) for the case $\bfr_1=\bfr_2=\bfr_3\,
,\,\bfk=0$, we substitute the ansatz $\eta_r\Omega^a_r$ into the
projected equation of (\ref{I15.1}). For brevity we do not
exhibit the corresponding calculations, rather we give the final
result. Combining the $\eta_r$ into a vector $\vec{\eta}$, we
get a homogeneous equation with the structure $A\, \vec{\eta}=0$, where the
matrix $A$ is given by
\begin{eqnarray*}
A&:=&\left(\begin{array}{ccc} \omega p+m+2/3\,\mu& \omega p+m+2/3\,\mu&0\\
2m+2/3\,\mu& \omega p-m-2/3\,\mu& 4m\\
-1/3\,\mu&1/3\,\mu&\omega p +m\end{array}\right) \\
\mu&:=&g\, s(0) \quad ,
\end{eqnarray*}
and $p\in \{ 1,-1\}$ is the parity of the three-particle states.
 From $\det A=0$ we get for the energy eigenvalues $\omega_i$:
\begin{eqnarray*}
\omega_1 &=& -p (m+2/3 \,\mu ) \\
\omega_2 &=& p\left[ 2/3 \sqrt{9m^2+12m\,\mu +\mu^2}+(m+2/3 \,\mu
)\right] \\
\omega_3 &=&- p\left[ 2/3 \sqrt{9m^2+12m\,\mu -\mu^2}+(m+2/3 \,\mu
)\right]
\end{eqnarray*}
Due to $\bfk =0$, the eigenvalues $\omega_i$ are the masses of
the bound states. They are given as a function of the coupling
constant in figure 1, where the region of the coupling  \\ \\[-2.5mm]
\begin{minipage}[b]{2in}
constant $g\sim \mu$ has been chosen to yield
$|\omega_i|<3m$. Furthermore, we see that there is a
region of $g$ which yields {\sl three} different eigenvalues $\omega_i$,
corresponding to {\sl three} linear independent vectors $\vec{\eta}$.
But we also recognize that in spite of
$|\omega_i|<3m$, the mass scale of the bound states
coincide with\hfill the\hfill mass\hfill scale\hfill of\hfill the
elementary fermions.
\end{minipage}
\hfill
\begin{minipage}[t]{3.7in}
\setlength{\unitlength}{0.240900pt}
\ifx\plotpoint\undefined\newsavebox{\plotpoint}\fi
\sbox{\plotpoint}{\rule[-0.200pt]{0.400pt}{0.400pt}}%
\begin{picture}(1125,765)(0,0)
\font\gnuplot=cmr10 at 10pt
\gnuplot
\sbox{\plotpoint}{\rule[-0.200pt]{0.400pt}{0.400pt}}%
\put(220.0,113.0){\rule[-0.200pt]{202.597pt}{0.400pt}}
\put(220.0,113.0){\rule[-0.200pt]{4.818pt}{0.400pt}}
\put(198,113){\makebox(0,0)[r]{0}}
\put(1041.0,113.0){\rule[-0.200pt]{4.818pt}{0.400pt}}
\put(220.0,186.0){\rule[-0.200pt]{4.818pt}{0.400pt}}
\put(198,186){\makebox(0,0)[r]{0.2}}
\put(1041.0,186.0){\rule[-0.200pt]{4.818pt}{0.400pt}}
\put(220.0,259.0){\rule[-0.200pt]{4.818pt}{0.400pt}}
\put(198,259){\makebox(0,0)[r]{0.4}}
\put(1041.0,259.0){\rule[-0.200pt]{4.818pt}{0.400pt}}
\put(220.0,332.0){\rule[-0.200pt]{4.818pt}{0.400pt}}
\put(198,332){\makebox(0,0)[r]{0.6}}
\put(1041.0,332.0){\rule[-0.200pt]{4.818pt}{0.400pt}}
\put(220.0,405.0){\rule[-0.200pt]{4.818pt}{0.400pt}}
\put(198,405){\makebox(0,0)[r]{0.8}}
\put(1041.0,405.0){\rule[-0.200pt]{4.818pt}{0.400pt}}
\put(220.0,478.0){\rule[-0.200pt]{4.818pt}{0.400pt}}
\put(198,478){\makebox(0,0)[r]{1}}
\put(1041.0,478.0){\rule[-0.200pt]{4.818pt}{0.400pt}}
\put(220.0,551.0){\rule[-0.200pt]{4.818pt}{0.400pt}}
\put(198,551){\makebox(0,0)[r]{1.2}}
\put(1041.0,551.0){\rule[-0.200pt]{4.818pt}{0.400pt}}
\put(220.0,624.0){\rule[-0.200pt]{4.818pt}{0.400pt}}
\put(198,624){\makebox(0,0)[r]{1.4}}
\put(1041.0,624.0){\rule[-0.200pt]{4.818pt}{0.400pt}}
\put(220.0,697.0){\rule[-0.200pt]{4.818pt}{0.400pt}}
\put(198,697){\makebox(0,0)[r]{1.6}}
\put(1041.0,697.0){\rule[-0.200pt]{4.818pt}{0.400pt}}
\put(220.0,113.0){\rule[-0.200pt]{0.400pt}{4.818pt}}
\put(220,68){\makebox(0,0){-0.82}}
\put(220.0,677.0){\rule[-0.200pt]{0.400pt}{4.818pt}}
\put(319.0,113.0){\rule[-0.200pt]{0.400pt}{4.818pt}}
\put(319,68){\makebox(0,0){-0.8}}
\put(319.0,677.0){\rule[-0.200pt]{0.400pt}{4.818pt}}
\put(418.0,113.0){\rule[-0.200pt]{0.400pt}{4.818pt}}
\put(418,68){\makebox(0,0){-0.78}}
\put(418.0,677.0){\rule[-0.200pt]{0.400pt}{4.818pt}}
\put(517.0,113.0){\rule[-0.200pt]{0.400pt}{4.818pt}}
\put(517,68){\makebox(0,0){-0.76}}
\put(517.0,677.0){\rule[-0.200pt]{0.400pt}{4.818pt}}
\put(616.0,113.0){\rule[-0.200pt]{0.400pt}{4.818pt}}
\put(616,68){\makebox(0,0){-0.74}}
\put(616.0,677.0){\rule[-0.200pt]{0.400pt}{4.818pt}}
\put(715.0,113.0){\rule[-0.200pt]{0.400pt}{4.818pt}}
\put(715,68){\makebox(0,0){-0.72}}
\put(715.0,677.0){\rule[-0.200pt]{0.400pt}{4.818pt}}
\put(814.0,113.0){\rule[-0.200pt]{0.400pt}{4.818pt}}
\put(814,68){\makebox(0,0){-0.7}}
\put(814.0,677.0){\rule[-0.200pt]{0.400pt}{4.818pt}}
\put(913.0,113.0){\rule[-0.200pt]{0.400pt}{4.818pt}}
\put(913,68){\makebox(0,0){-0.68}}
\put(913.0,677.0){\rule[-0.200pt]{0.400pt}{4.818pt}}
\put(1012.0,113.0){\rule[-0.200pt]{0.400pt}{4.818pt}}
\put(1012,68){\makebox(0,0){-0.66}}
\put(1012.0,677.0){\rule[-0.200pt]{0.400pt}{4.818pt}}
\put(220.0,113.0){\rule[-0.200pt]{202.597pt}{0.400pt}}
\put(1061.0,113.0){\rule[-0.200pt]{0.400pt}{140.686pt}}
\put(220.0,697.0){\rule[-0.200pt]{202.597pt}{0.400pt}}
\put(45,405){\makebox(0,0){$|\frac{\omega}{m}|$}}
\put(640,23){\makebox(0,0){$\mu /m\, =\, g\, s(0)/m$}}
\put(640,742){\makebox(0,0){Masses of the Boundstates}}
\put(517,588){\makebox(0,0){Figure 1}}
\put(329,175){\makebox(0,0){$ |\omega_3\! |$}}
\put(418,314){\makebox(0,0){$ |\omega_1\! |$}}
\put(319,405){\makebox(0,0){$ |\omega_2\! |$}}
\put(220.0,113.0){\rule[-0.200pt]{0.400pt}{140.686pt}}
\put(305,308){\usebox{\plotpoint}}
\multiput(305.59,308.00)(0.488,1.022){13}{\rule{0.117pt}{0.900pt}}
\multiput(304.17,308.00)(8.000,14.132){2}{\rule{0.400pt}{0.450pt}}
\multiput(313.59,324.00)(0.489,0.669){15}{\rule{0.118pt}{0.633pt}}
\multiput(312.17,324.00)(9.000,10.685){2}{\rule{0.400pt}{0.317pt}}
\multiput(322.59,336.00)(0.488,0.626){13}{\rule{0.117pt}{0.600pt}}
\multiput(321.17,336.00)(8.000,8.755){2}{\rule{0.400pt}{0.300pt}}
\multiput(330.00,346.59)(0.560,0.488){13}{\rule{0.550pt}{0.117pt}}
\multiput(330.00,345.17)(7.858,8.000){2}{\rule{0.275pt}{0.400pt}}
\multiput(339.00,354.59)(0.494,0.488){13}{\rule{0.500pt}{0.117pt}}
\multiput(339.00,353.17)(6.962,8.000){2}{\rule{0.250pt}{0.400pt}}
\multiput(347.00,362.59)(0.645,0.485){11}{\rule{0.614pt}{0.117pt}}
\multiput(347.00,361.17)(7.725,7.000){2}{\rule{0.307pt}{0.400pt}}
\multiput(356.00,369.59)(0.671,0.482){9}{\rule{0.633pt}{0.116pt}}
\multiput(356.00,368.17)(6.685,6.000){2}{\rule{0.317pt}{0.400pt}}
\multiput(364.00,375.59)(0.762,0.482){9}{\rule{0.700pt}{0.116pt}}
\multiput(364.00,374.17)(7.547,6.000){2}{\rule{0.350pt}{0.400pt}}
\multiput(373.00,381.59)(0.671,0.482){9}{\rule{0.633pt}{0.116pt}}
\multiput(373.00,380.17)(6.685,6.000){2}{\rule{0.317pt}{0.400pt}}
\multiput(381.00,387.59)(0.762,0.482){9}{\rule{0.700pt}{0.116pt}}
\multiput(381.00,386.17)(7.547,6.000){2}{\rule{0.350pt}{0.400pt}}
\multiput(390.00,393.59)(0.821,0.477){7}{\rule{0.740pt}{0.115pt}}
\multiput(390.00,392.17)(6.464,5.000){2}{\rule{0.370pt}{0.400pt}}
\multiput(398.00,398.59)(0.933,0.477){7}{\rule{0.820pt}{0.115pt}}
\multiput(398.00,397.17)(7.298,5.000){2}{\rule{0.410pt}{0.400pt}}
\multiput(407.00,403.59)(0.821,0.477){7}{\rule{0.740pt}{0.115pt}}
\multiput(407.00,402.17)(6.464,5.000){2}{\rule{0.370pt}{0.400pt}}
\multiput(415.00,408.59)(0.933,0.477){7}{\rule{0.820pt}{0.115pt}}
\multiput(415.00,407.17)(7.298,5.000){2}{\rule{0.410pt}{0.400pt}}
\multiput(424.00,413.60)(1.066,0.468){5}{\rule{0.900pt}{0.113pt}}
\multiput(424.00,412.17)(6.132,4.000){2}{\rule{0.450pt}{0.400pt}}
\multiput(432.00,417.59)(0.933,0.477){7}{\rule{0.820pt}{0.115pt}}
\multiput(432.00,416.17)(7.298,5.000){2}{\rule{0.410pt}{0.400pt}}
\multiput(441.00,422.60)(1.066,0.468){5}{\rule{0.900pt}{0.113pt}}
\multiput(441.00,421.17)(6.132,4.000){2}{\rule{0.450pt}{0.400pt}}
\multiput(449.00,426.59)(0.933,0.477){7}{\rule{0.820pt}{0.115pt}}
\multiput(449.00,425.17)(7.298,5.000){2}{\rule{0.410pt}{0.400pt}}
\multiput(458.00,431.60)(1.066,0.468){5}{\rule{0.900pt}{0.113pt}}
\multiput(458.00,430.17)(6.132,4.000){2}{\rule{0.450pt}{0.400pt}}
\multiput(466.00,435.60)(1.212,0.468){5}{\rule{1.000pt}{0.113pt}}
\multiput(466.00,434.17)(6.924,4.000){2}{\rule{0.500pt}{0.400pt}}
\multiput(475.00,439.60)(1.066,0.468){5}{\rule{0.900pt}{0.113pt}}
\multiput(475.00,438.17)(6.132,4.000){2}{\rule{0.450pt}{0.400pt}}
\multiput(483.00,443.60)(1.212,0.468){5}{\rule{1.000pt}{0.113pt}}
\multiput(483.00,442.17)(6.924,4.000){2}{\rule{0.500pt}{0.400pt}}
\multiput(492.00,447.61)(1.579,0.447){3}{\rule{1.167pt}{0.108pt}}
\multiput(492.00,446.17)(5.579,3.000){2}{\rule{0.583pt}{0.400pt}}
\multiput(500.00,450.60)(1.212,0.468){5}{\rule{1.000pt}{0.113pt}}
\multiput(500.00,449.17)(6.924,4.000){2}{\rule{0.500pt}{0.400pt}}
\multiput(509.00,454.60)(1.066,0.468){5}{\rule{0.900pt}{0.113pt}}
\multiput(509.00,453.17)(6.132,4.000){2}{\rule{0.450pt}{0.400pt}}
\multiput(517.00,458.60)(1.212,0.468){5}{\rule{1.000pt}{0.113pt}}
\multiput(517.00,457.17)(6.924,4.000){2}{\rule{0.500pt}{0.400pt}}
\multiput(526.00,462.61)(1.579,0.447){3}{\rule{1.167pt}{0.108pt}}
\multiput(526.00,461.17)(5.579,3.000){2}{\rule{0.583pt}{0.400pt}}
\multiput(534.00,465.60)(1.212,0.468){5}{\rule{1.000pt}{0.113pt}}
\multiput(534.00,464.17)(6.924,4.000){2}{\rule{0.500pt}{0.400pt}}
\multiput(543.00,469.61)(1.579,0.447){3}{\rule{1.167pt}{0.108pt}}
\multiput(543.00,468.17)(5.579,3.000){2}{\rule{0.583pt}{0.400pt}}
\multiput(551.00,472.61)(1.802,0.447){3}{\rule{1.300pt}{0.108pt}}
\multiput(551.00,471.17)(6.302,3.000){2}{\rule{0.650pt}{0.400pt}}
\multiput(560.00,475.60)(1.066,0.468){5}{\rule{0.900pt}{0.113pt}}
\multiput(560.00,474.17)(6.132,4.000){2}{\rule{0.450pt}{0.400pt}}
\multiput(568.00,479.61)(1.802,0.447){3}{\rule{1.300pt}{0.108pt}}
\multiput(568.00,478.17)(6.302,3.000){2}{\rule{0.650pt}{0.400pt}}
\multiput(577.00,482.61)(1.579,0.447){3}{\rule{1.167pt}{0.108pt}}
\multiput(577.00,481.17)(5.579,3.000){2}{\rule{0.583pt}{0.400pt}}
\multiput(585.00,485.60)(1.212,0.468){5}{\rule{1.000pt}{0.113pt}}
\multiput(585.00,484.17)(6.924,4.000){2}{\rule{0.500pt}{0.400pt}}
\multiput(594.00,489.61)(1.579,0.447){3}{\rule{1.167pt}{0.108pt}}
\multiput(594.00,488.17)(5.579,3.000){2}{\rule{0.583pt}{0.400pt}}
\multiput(602.00,492.61)(1.802,0.447){3}{\rule{1.300pt}{0.108pt}}
\multiput(602.00,491.17)(6.302,3.000){2}{\rule{0.650pt}{0.400pt}}
\multiput(611.00,495.61)(1.579,0.447){3}{\rule{1.167pt}{0.108pt}}
\multiput(611.00,494.17)(5.579,3.000){2}{\rule{0.583pt}{0.400pt}}
\multiput(619.00,498.61)(1.802,0.447){3}{\rule{1.300pt}{0.108pt}}
\multiput(619.00,497.17)(6.302,3.000){2}{\rule{0.650pt}{0.400pt}}
\multiput(628.00,501.61)(1.579,0.447){3}{\rule{1.167pt}{0.108pt}}
\multiput(628.00,500.17)(5.579,3.000){2}{\rule{0.583pt}{0.400pt}}
\multiput(636.00,504.61)(1.802,0.447){3}{\rule{1.300pt}{0.108pt}}
\multiput(636.00,503.17)(6.302,3.000){2}{\rule{0.650pt}{0.400pt}}
\multiput(645.00,507.61)(1.579,0.447){3}{\rule{1.167pt}{0.108pt}}
\multiput(645.00,506.17)(5.579,3.000){2}{\rule{0.583pt}{0.400pt}}
\multiput(653.00,510.61)(1.802,0.447){3}{\rule{1.300pt}{0.108pt}}
\multiput(653.00,509.17)(6.302,3.000){2}{\rule{0.650pt}{0.400pt}}
\multiput(662.00,513.61)(1.579,0.447){3}{\rule{1.167pt}{0.108pt}}
\multiput(662.00,512.17)(5.579,3.000){2}{\rule{0.583pt}{0.400pt}}
\multiput(670.00,516.61)(1.802,0.447){3}{\rule{1.300pt}{0.108pt}}
\multiput(670.00,515.17)(6.302,3.000){2}{\rule{0.650pt}{0.400pt}}
\multiput(679.00,519.61)(1.579,0.447){3}{\rule{1.167pt}{0.108pt}}
\multiput(679.00,518.17)(5.579,3.000){2}{\rule{0.583pt}{0.400pt}}
\multiput(687.00,522.61)(1.802,0.447){3}{\rule{1.300pt}{0.108pt}}
\multiput(687.00,521.17)(6.302,3.000){2}{\rule{0.650pt}{0.400pt}}
\put(696,525.17){\rule{1.700pt}{0.400pt}}
\multiput(696.00,524.17)(4.472,2.000){2}{\rule{0.850pt}{0.400pt}}
\multiput(704.00,527.61)(1.802,0.447){3}{\rule{1.300pt}{0.108pt}}
\multiput(704.00,526.17)(6.302,3.000){2}{\rule{0.650pt}{0.400pt}}
\multiput(713.00,530.61)(1.579,0.447){3}{\rule{1.167pt}{0.108pt}}
\multiput(713.00,529.17)(5.579,3.000){2}{\rule{0.583pt}{0.400pt}}
\multiput(721.00,533.61)(1.802,0.447){3}{\rule{1.300pt}{0.108pt}}
\multiput(721.00,532.17)(6.302,3.000){2}{\rule{0.650pt}{0.400pt}}
\put(730,536.17){\rule{1.700pt}{0.400pt}}
\multiput(730.00,535.17)(4.472,2.000){2}{\rule{0.850pt}{0.400pt}}
\multiput(738.00,538.61)(1.802,0.447){3}{\rule{1.300pt}{0.108pt}}
\multiput(738.00,537.17)(6.302,3.000){2}{\rule{0.650pt}{0.400pt}}
\multiput(747.00,541.61)(1.579,0.447){3}{\rule{1.167pt}{0.108pt}}
\multiput(747.00,540.17)(5.579,3.000){2}{\rule{0.583pt}{0.400pt}}
\put(755,544.17){\rule{1.900pt}{0.400pt}}
\multiput(755.00,543.17)(5.056,2.000){2}{\rule{0.950pt}{0.400pt}}
\multiput(764.00,546.61)(1.579,0.447){3}{\rule{1.167pt}{0.108pt}}
\multiput(764.00,545.17)(5.579,3.000){2}{\rule{0.583pt}{0.400pt}}
\multiput(772.00,549.61)(1.802,0.447){3}{\rule{1.300pt}{0.108pt}}
\multiput(772.00,548.17)(6.302,3.000){2}{\rule{0.650pt}{0.400pt}}
\put(781,552.17){\rule{1.700pt}{0.400pt}}
\multiput(781.00,551.17)(4.472,2.000){2}{\rule{0.850pt}{0.400pt}}
\multiput(789.00,554.61)(1.802,0.447){3}{\rule{1.300pt}{0.108pt}}
\multiput(789.00,553.17)(6.302,3.000){2}{\rule{0.650pt}{0.400pt}}
\put(798,557.17){\rule{1.700pt}{0.400pt}}
\multiput(798.00,556.17)(4.472,2.000){2}{\rule{0.850pt}{0.400pt}}
\multiput(806.00,559.61)(1.802,0.447){3}{\rule{1.300pt}{0.108pt}}
\multiput(806.00,558.17)(6.302,3.000){2}{\rule{0.650pt}{0.400pt}}
\multiput(815.00,562.61)(1.579,0.447){3}{\rule{1.167pt}{0.108pt}}
\multiput(815.00,561.17)(5.579,3.000){2}{\rule{0.583pt}{0.400pt}}
\put(823,565.17){\rule{1.900pt}{0.400pt}}
\multiput(823.00,564.17)(5.056,2.000){2}{\rule{0.950pt}{0.400pt}}
\multiput(832.00,567.61)(1.579,0.447){3}{\rule{1.167pt}{0.108pt}}
\multiput(832.00,566.17)(5.579,3.000){2}{\rule{0.583pt}{0.400pt}}
\put(840,570.17){\rule{1.900pt}{0.400pt}}
\multiput(840.00,569.17)(5.056,2.000){2}{\rule{0.950pt}{0.400pt}}
\put(849,572.17){\rule{1.700pt}{0.400pt}}
\multiput(849.00,571.17)(4.472,2.000){2}{\rule{0.850pt}{0.400pt}}
\multiput(857.00,574.61)(1.802,0.447){3}{\rule{1.300pt}{0.108pt}}
\multiput(857.00,573.17)(6.302,3.000){2}{\rule{0.650pt}{0.400pt}}
\put(866,577.17){\rule{1.700pt}{0.400pt}}
\multiput(866.00,576.17)(4.472,2.000){2}{\rule{0.850pt}{0.400pt}}
\multiput(874.00,579.61)(1.802,0.447){3}{\rule{1.300pt}{0.108pt}}
\multiput(874.00,578.17)(6.302,3.000){2}{\rule{0.650pt}{0.400pt}}
\put(883,582.17){\rule{1.700pt}{0.400pt}}
\multiput(883.00,581.17)(4.472,2.000){2}{\rule{0.850pt}{0.400pt}}
\multiput(891.00,584.61)(1.802,0.447){3}{\rule{1.300pt}{0.108pt}}
\multiput(891.00,583.17)(6.302,3.000){2}{\rule{0.650pt}{0.400pt}}
\put(900,587.17){\rule{1.700pt}{0.400pt}}
\multiput(900.00,586.17)(4.472,2.000){2}{\rule{0.850pt}{0.400pt}}
\put(908,589.17){\rule{1.900pt}{0.400pt}}
\multiput(908.00,588.17)(5.056,2.000){2}{\rule{0.950pt}{0.400pt}}
\multiput(917.00,591.61)(1.579,0.447){3}{\rule{1.167pt}{0.108pt}}
\multiput(917.00,590.17)(5.579,3.000){2}{\rule{0.583pt}{0.400pt}}
\put(925,594.17){\rule{1.900pt}{0.400pt}}
\multiput(925.00,593.17)(5.056,2.000){2}{\rule{0.950pt}{0.400pt}}
\put(934,596.17){\rule{1.700pt}{0.400pt}}
\multiput(934.00,595.17)(4.472,2.000){2}{\rule{0.850pt}{0.400pt}}
\multiput(942.00,598.61)(1.802,0.447){3}{\rule{1.300pt}{0.108pt}}
\multiput(942.00,597.17)(6.302,3.000){2}{\rule{0.650pt}{0.400pt}}
\put(951,601.17){\rule{1.700pt}{0.400pt}}
\multiput(951.00,600.17)(4.472,2.000){2}{\rule{0.850pt}{0.400pt}}
\put(959,603.17){\rule{1.900pt}{0.400pt}}
\multiput(959.00,602.17)(5.056,2.000){2}{\rule{0.950pt}{0.400pt}}
\multiput(968.00,605.61)(1.579,0.447){3}{\rule{1.167pt}{0.108pt}}
\multiput(968.00,604.17)(5.579,3.000){2}{\rule{0.583pt}{0.400pt}}
\put(976,608.17){\rule{1.900pt}{0.400pt}}
\multiput(976.00,607.17)(5.056,2.000){2}{\rule{0.950pt}{0.400pt}}
\put(985,610.17){\rule{1.700pt}{0.400pt}}
\multiput(985.00,609.17)(4.472,2.000){2}{\rule{0.850pt}{0.400pt}}
\put(993,612.17){\rule{1.900pt}{0.400pt}}
\multiput(993.00,611.17)(5.056,2.000){2}{\rule{0.950pt}{0.400pt}}
\multiput(1002.00,614.61)(1.579,0.447){3}{\rule{1.167pt}{0.108pt}}
\multiput(1002.00,613.17)(5.579,3.000){2}{\rule{0.583pt}{0.400pt}}
\put(1010,617.17){\rule{1.900pt}{0.400pt}}
\multiput(1010.00,616.17)(5.056,2.000){2}{\rule{0.950pt}{0.400pt}}
\put(1019,619.17){\rule{1.700pt}{0.400pt}}
\multiput(1019.00,618.17)(4.472,2.000){2}{\rule{0.850pt}{0.400pt}}
\put(1027,621.17){\rule{1.900pt}{0.400pt}}
\multiput(1027.00,620.17)(5.056,2.000){2}{\rule{0.950pt}{0.400pt}}
\put(1036,623.17){\rule{1.700pt}{0.400pt}}
\multiput(1036.00,622.17)(4.472,2.000){2}{\rule{0.850pt}{0.400pt}}
\multiput(1044.00,625.61)(1.802,0.447){3}{\rule{1.300pt}{0.108pt}}
\multiput(1044.00,624.17)(6.302,3.000){2}{\rule{0.650pt}{0.400pt}}
\put(1053,628.17){\rule{1.700pt}{0.400pt}}
\multiput(1053.00,627.17)(4.472,2.000){2}{\rule{0.850pt}{0.400pt}}
\put(305,258){\usebox{\plotpoint}}
\multiput(305.59,254.26)(0.488,-1.022){13}{\rule{0.117pt}{0.900pt}}
\multiput(304.17,256.13)(8.000,-14.132){2}{\rule{0.400pt}{0.450pt}}
\multiput(313.59,239.56)(0.489,-0.611){15}{\rule{0.118pt}{0.589pt}}
\multiput(312.17,240.78)(9.000,-9.778){2}{\rule{0.400pt}{0.294pt}}
\multiput(322.59,228.72)(0.488,-0.560){13}{\rule{0.117pt}{0.550pt}}
\multiput(321.17,229.86)(8.000,-7.858){2}{\rule{0.400pt}{0.275pt}}
\multiput(330.00,220.93)(0.645,-0.485){11}{\rule{0.614pt}{0.117pt}}
\multiput(330.00,221.17)(7.725,-7.000){2}{\rule{0.307pt}{0.400pt}}
\multiput(339.00,213.93)(0.569,-0.485){11}{\rule{0.557pt}{0.117pt}}
\multiput(339.00,214.17)(6.844,-7.000){2}{\rule{0.279pt}{0.400pt}}
\multiput(347.00,206.93)(0.762,-0.482){9}{\rule{0.700pt}{0.116pt}}
\multiput(347.00,207.17)(7.547,-6.000){2}{\rule{0.350pt}{0.400pt}}
\multiput(356.00,200.93)(0.671,-0.482){9}{\rule{0.633pt}{0.116pt}}
\multiput(356.00,201.17)(6.685,-6.000){2}{\rule{0.317pt}{0.400pt}}
\multiput(364.00,194.93)(0.933,-0.477){7}{\rule{0.820pt}{0.115pt}}
\multiput(364.00,195.17)(7.298,-5.000){2}{\rule{0.410pt}{0.400pt}}
\multiput(373.00,189.93)(0.821,-0.477){7}{\rule{0.740pt}{0.115pt}}
\multiput(373.00,190.17)(6.464,-5.000){2}{\rule{0.370pt}{0.400pt}}
\multiput(381.00,184.93)(0.933,-0.477){7}{\rule{0.820pt}{0.115pt}}
\multiput(381.00,185.17)(7.298,-5.000){2}{\rule{0.410pt}{0.400pt}}
\multiput(390.00,179.93)(0.821,-0.477){7}{\rule{0.740pt}{0.115pt}}
\multiput(390.00,180.17)(6.464,-5.000){2}{\rule{0.370pt}{0.400pt}}
\multiput(398.00,174.94)(1.212,-0.468){5}{\rule{1.000pt}{0.113pt}}
\multiput(398.00,175.17)(6.924,-4.000){2}{\rule{0.500pt}{0.400pt}}
\multiput(407.00,170.94)(1.066,-0.468){5}{\rule{0.900pt}{0.113pt}}
\multiput(407.00,171.17)(6.132,-4.000){2}{\rule{0.450pt}{0.400pt}}
\multiput(415.00,166.94)(1.212,-0.468){5}{\rule{1.000pt}{0.113pt}}
\multiput(415.00,167.17)(6.924,-4.000){2}{\rule{0.500pt}{0.400pt}}
\multiput(424.00,162.94)(1.066,-0.468){5}{\rule{0.900pt}{0.113pt}}
\multiput(424.00,163.17)(6.132,-4.000){2}{\rule{0.450pt}{0.400pt}}
\multiput(432.00,158.95)(1.802,-0.447){3}{\rule{1.300pt}{0.108pt}}
\multiput(432.00,159.17)(6.302,-3.000){2}{\rule{0.650pt}{0.400pt}}
\multiput(441.00,155.94)(1.066,-0.468){5}{\rule{0.900pt}{0.113pt}}
\multiput(441.00,156.17)(6.132,-4.000){2}{\rule{0.450pt}{0.400pt}}
\multiput(449.00,151.95)(1.802,-0.447){3}{\rule{1.300pt}{0.108pt}}
\multiput(449.00,152.17)(6.302,-3.000){2}{\rule{0.650pt}{0.400pt}}
\multiput(458.00,148.94)(1.066,-0.468){5}{\rule{0.900pt}{0.113pt}}
\multiput(458.00,149.17)(6.132,-4.000){2}{\rule{0.450pt}{0.400pt}}
\multiput(466.00,144.95)(1.802,-0.447){3}{\rule{1.300pt}{0.108pt}}
\multiput(466.00,145.17)(6.302,-3.000){2}{\rule{0.650pt}{0.400pt}}
\multiput(475.00,141.95)(1.579,-0.447){3}{\rule{1.167pt}{0.108pt}}
\multiput(475.00,142.17)(5.579,-3.000){2}{\rule{0.583pt}{0.400pt}}
\multiput(483.00,138.95)(1.802,-0.447){3}{\rule{1.300pt}{0.108pt}}
\multiput(483.00,139.17)(6.302,-3.000){2}{\rule{0.650pt}{0.400pt}}
\multiput(492.00,135.95)(1.579,-0.447){3}{\rule{1.167pt}{0.108pt}}
\multiput(492.00,136.17)(5.579,-3.000){2}{\rule{0.583pt}{0.400pt}}
\multiput(500.00,132.95)(1.802,-0.447){3}{\rule{1.300pt}{0.108pt}}
\multiput(500.00,133.17)(6.302,-3.000){2}{\rule{0.650pt}{0.400pt}}
\multiput(509.00,129.95)(1.579,-0.447){3}{\rule{1.167pt}{0.108pt}}
\multiput(509.00,130.17)(5.579,-3.000){2}{\rule{0.583pt}{0.400pt}}
\put(517,126.17){\rule{1.900pt}{0.400pt}}
\multiput(517.00,127.17)(5.056,-2.000){2}{\rule{0.950pt}{0.400pt}}
\multiput(526.00,124.95)(1.579,-0.447){3}{\rule{1.167pt}{0.108pt}}
\multiput(526.00,125.17)(5.579,-3.000){2}{\rule{0.583pt}{0.400pt}}
\multiput(534.00,121.95)(1.802,-0.447){3}{\rule{1.300pt}{0.108pt}}
\multiput(534.00,122.17)(6.302,-3.000){2}{\rule{0.650pt}{0.400pt}}
\multiput(543.00,118.95)(1.579,-0.447){3}{\rule{1.167pt}{0.108pt}}
\multiput(543.00,119.17)(5.579,-3.000){2}{\rule{0.583pt}{0.400pt}}
\put(551,115.17){\rule{1.900pt}{0.400pt}}
\multiput(551.00,116.17)(5.056,-2.000){2}{\rule{0.950pt}{0.400pt}}
\put(560,113.67){\rule{1.927pt}{0.400pt}}
\multiput(560.00,114.17)(4.000,-1.000){2}{\rule{0.964pt}{0.400pt}}
\put(568,114.17){\rule{1.900pt}{0.400pt}}
\multiput(568.00,113.17)(5.056,2.000){2}{\rule{0.950pt}{0.400pt}}
\put(577,116.17){\rule{1.700pt}{0.400pt}}
\multiput(577.00,115.17)(4.472,2.000){2}{\rule{0.850pt}{0.400pt}}
\multiput(585.00,118.61)(1.802,0.447){3}{\rule{1.300pt}{0.108pt}}
\multiput(585.00,117.17)(6.302,3.000){2}{\rule{0.650pt}{0.400pt}}
\put(594,121.17){\rule{1.700pt}{0.400pt}}
\multiput(594.00,120.17)(4.472,2.000){2}{\rule{0.850pt}{0.400pt}}
\multiput(602.00,123.61)(1.802,0.447){3}{\rule{1.300pt}{0.108pt}}
\multiput(602.00,122.17)(6.302,3.000){2}{\rule{0.650pt}{0.400pt}}
\put(611,126.17){\rule{1.700pt}{0.400pt}}
\multiput(611.00,125.17)(4.472,2.000){2}{\rule{0.850pt}{0.400pt}}
\put(619,128.17){\rule{1.900pt}{0.400pt}}
\multiput(619.00,127.17)(5.056,2.000){2}{\rule{0.950pt}{0.400pt}}
\put(628,130.17){\rule{1.700pt}{0.400pt}}
\multiput(628.00,129.17)(4.472,2.000){2}{\rule{0.850pt}{0.400pt}}
\put(636,132.17){\rule{1.900pt}{0.400pt}}
\multiput(636.00,131.17)(5.056,2.000){2}{\rule{0.950pt}{0.400pt}}
\multiput(645.00,134.61)(1.579,0.447){3}{\rule{1.167pt}{0.108pt}}
\multiput(645.00,133.17)(5.579,3.000){2}{\rule{0.583pt}{0.400pt}}
\put(653,137.17){\rule{1.900pt}{0.400pt}}
\multiput(653.00,136.17)(5.056,2.000){2}{\rule{0.950pt}{0.400pt}}
\put(662,139.17){\rule{1.700pt}{0.400pt}}
\multiput(662.00,138.17)(4.472,2.000){2}{\rule{0.850pt}{0.400pt}}
\put(670,141.17){\rule{1.900pt}{0.400pt}}
\multiput(670.00,140.17)(5.056,2.000){2}{\rule{0.950pt}{0.400pt}}
\put(679,143.17){\rule{1.700pt}{0.400pt}}
\multiput(679.00,142.17)(4.472,2.000){2}{\rule{0.850pt}{0.400pt}}
\put(687,145.17){\rule{1.900pt}{0.400pt}}
\multiput(687.00,144.17)(5.056,2.000){2}{\rule{0.950pt}{0.400pt}}
\put(696,147.17){\rule{1.700pt}{0.400pt}}
\multiput(696.00,146.17)(4.472,2.000){2}{\rule{0.850pt}{0.400pt}}
\put(704,149.17){\rule{1.900pt}{0.400pt}}
\multiput(704.00,148.17)(5.056,2.000){2}{\rule{0.950pt}{0.400pt}}
\put(713,151.17){\rule{1.700pt}{0.400pt}}
\multiput(713.00,150.17)(4.472,2.000){2}{\rule{0.850pt}{0.400pt}}
\put(721,153.17){\rule{1.900pt}{0.400pt}}
\multiput(721.00,152.17)(5.056,2.000){2}{\rule{0.950pt}{0.400pt}}
\put(730,155.17){\rule{1.700pt}{0.400pt}}
\multiput(730.00,154.17)(4.472,2.000){2}{\rule{0.850pt}{0.400pt}}
\put(738,156.67){\rule{2.168pt}{0.400pt}}
\multiput(738.00,156.17)(4.500,1.000){2}{\rule{1.084pt}{0.400pt}}
\put(747,158.17){\rule{1.700pt}{0.400pt}}
\multiput(747.00,157.17)(4.472,2.000){2}{\rule{0.850pt}{0.400pt}}
\put(755,160.17){\rule{1.900pt}{0.400pt}}
\multiput(755.00,159.17)(5.056,2.000){2}{\rule{0.950pt}{0.400pt}}
\put(764,162.17){\rule{1.700pt}{0.400pt}}
\multiput(764.00,161.17)(4.472,2.000){2}{\rule{0.850pt}{0.400pt}}
\put(772,164.17){\rule{1.900pt}{0.400pt}}
\multiput(772.00,163.17)(5.056,2.000){2}{\rule{0.950pt}{0.400pt}}
\put(781,165.67){\rule{1.927pt}{0.400pt}}
\multiput(781.00,165.17)(4.000,1.000){2}{\rule{0.964pt}{0.400pt}}
\put(789,167.17){\rule{1.900pt}{0.400pt}}
\multiput(789.00,166.17)(5.056,2.000){2}{\rule{0.950pt}{0.400pt}}
\put(798,169.17){\rule{1.700pt}{0.400pt}}
\multiput(798.00,168.17)(4.472,2.000){2}{\rule{0.850pt}{0.400pt}}
\put(806,171.17){\rule{1.900pt}{0.400pt}}
\multiput(806.00,170.17)(5.056,2.000){2}{\rule{0.950pt}{0.400pt}}
\put(815,172.67){\rule{1.927pt}{0.400pt}}
\multiput(815.00,172.17)(4.000,1.000){2}{\rule{0.964pt}{0.400pt}}
\put(823,174.17){\rule{1.900pt}{0.400pt}}
\multiput(823.00,173.17)(5.056,2.000){2}{\rule{0.950pt}{0.400pt}}
\put(832,176.17){\rule{1.700pt}{0.400pt}}
\multiput(832.00,175.17)(4.472,2.000){2}{\rule{0.850pt}{0.400pt}}
\put(840,177.67){\rule{2.168pt}{0.400pt}}
\multiput(840.00,177.17)(4.500,1.000){2}{\rule{1.084pt}{0.400pt}}
\put(849,179.17){\rule{1.700pt}{0.400pt}}
\multiput(849.00,178.17)(4.472,2.000){2}{\rule{0.850pt}{0.400pt}}
\put(857,180.67){\rule{2.168pt}{0.400pt}}
\multiput(857.00,180.17)(4.500,1.000){2}{\rule{1.084pt}{0.400pt}}
\put(866,182.17){\rule{1.700pt}{0.400pt}}
\multiput(866.00,181.17)(4.472,2.000){2}{\rule{0.850pt}{0.400pt}}
\put(874,184.17){\rule{1.900pt}{0.400pt}}
\multiput(874.00,183.17)(5.056,2.000){2}{\rule{0.950pt}{0.400pt}}
\put(883,185.67){\rule{1.927pt}{0.400pt}}
\multiput(883.00,185.17)(4.000,1.000){2}{\rule{0.964pt}{0.400pt}}
\put(891,187.17){\rule{1.900pt}{0.400pt}}
\multiput(891.00,186.17)(5.056,2.000){2}{\rule{0.950pt}{0.400pt}}
\put(900,188.67){\rule{1.927pt}{0.400pt}}
\multiput(900.00,188.17)(4.000,1.000){2}{\rule{0.964pt}{0.400pt}}
\put(908,190.17){\rule{1.900pt}{0.400pt}}
\multiput(908.00,189.17)(5.056,2.000){2}{\rule{0.950pt}{0.400pt}}
\put(917,191.67){\rule{1.927pt}{0.400pt}}
\multiput(917.00,191.17)(4.000,1.000){2}{\rule{0.964pt}{0.400pt}}
\put(925,193.17){\rule{1.900pt}{0.400pt}}
\multiput(925.00,192.17)(5.056,2.000){2}{\rule{0.950pt}{0.400pt}}
\put(934,194.67){\rule{1.927pt}{0.400pt}}
\multiput(934.00,194.17)(4.000,1.000){2}{\rule{0.964pt}{0.400pt}}
\put(942,196.17){\rule{1.900pt}{0.400pt}}
\multiput(942.00,195.17)(5.056,2.000){2}{\rule{0.950pt}{0.400pt}}
\put(951,197.67){\rule{1.927pt}{0.400pt}}
\multiput(951.00,197.17)(4.000,1.000){2}{\rule{0.964pt}{0.400pt}}
\put(959,199.17){\rule{1.900pt}{0.400pt}}
\multiput(959.00,198.17)(5.056,2.000){2}{\rule{0.950pt}{0.400pt}}
\put(968,200.67){\rule{1.927pt}{0.400pt}}
\multiput(968.00,200.17)(4.000,1.000){2}{\rule{0.964pt}{0.400pt}}
\put(976,202.17){\rule{1.900pt}{0.400pt}}
\multiput(976.00,201.17)(5.056,2.000){2}{\rule{0.950pt}{0.400pt}}
\put(985,203.67){\rule{1.927pt}{0.400pt}}
\multiput(985.00,203.17)(4.000,1.000){2}{\rule{0.964pt}{0.400pt}}
\put(993,205.17){\rule{1.900pt}{0.400pt}}
\multiput(993.00,204.17)(5.056,2.000){2}{\rule{0.950pt}{0.400pt}}
\put(1002,206.67){\rule{1.927pt}{0.400pt}}
\multiput(1002.00,206.17)(4.000,1.000){2}{\rule{0.964pt}{0.400pt}}
\put(1010,207.67){\rule{2.168pt}{0.400pt}}
\multiput(1010.00,207.17)(4.500,1.000){2}{\rule{1.084pt}{0.400pt}}
\put(1019,209.17){\rule{1.700pt}{0.400pt}}
\multiput(1019.00,208.17)(4.472,2.000){2}{\rule{0.850pt}{0.400pt}}
\put(1027,210.67){\rule{2.168pt}{0.400pt}}
\multiput(1027.00,210.17)(4.500,1.000){2}{\rule{1.084pt}{0.400pt}}
\put(1036,211.67){\rule{1.927pt}{0.400pt}}
\multiput(1036.00,211.17)(4.000,1.000){2}{\rule{0.964pt}{0.400pt}}
\put(1044,213.17){\rule{1.900pt}{0.400pt}}
\multiput(1044.00,212.17)(5.056,2.000){2}{\rule{0.950pt}{0.400pt}}
\put(1053,214.67){\rule{1.927pt}{0.400pt}}
\multiput(1053.00,214.17)(4.000,1.000){2}{\rule{0.964pt}{0.400pt}}
\put(220,278){\usebox{\plotpoint}}
\put(220,277.67){\rule{1.927pt}{0.400pt}}
\multiput(220.00,277.17)(4.000,1.000){2}{\rule{0.964pt}{0.400pt}}
\put(237,278.67){\rule{1.927pt}{0.400pt}}
\multiput(237.00,278.17)(4.000,1.000){2}{\rule{0.964pt}{0.400pt}}
\put(228.0,279.0){\rule[-0.200pt]{2.168pt}{0.400pt}}
\put(254,279.67){\rule{1.927pt}{0.400pt}}
\multiput(254.00,279.17)(4.000,1.000){2}{\rule{0.964pt}{0.400pt}}
\put(245.0,280.0){\rule[-0.200pt]{2.168pt}{0.400pt}}
\put(279,280.67){\rule{2.168pt}{0.400pt}}
\multiput(279.00,280.17)(4.500,1.000){2}{\rule{1.084pt}{0.400pt}}
\put(262.0,281.0){\rule[-0.200pt]{4.095pt}{0.400pt}}
\put(296,281.67){\rule{2.168pt}{0.400pt}}
\multiput(296.00,281.17)(4.500,1.000){2}{\rule{1.084pt}{0.400pt}}
\put(288.0,282.0){\rule[-0.200pt]{1.927pt}{0.400pt}}
\put(322,282.67){\rule{1.927pt}{0.400pt}}
\multiput(322.00,282.17)(4.000,1.000){2}{\rule{0.964pt}{0.400pt}}
\put(305.0,283.0){\rule[-0.200pt]{4.095pt}{0.400pt}}
\put(339,283.67){\rule{1.927pt}{0.400pt}}
\multiput(339.00,283.17)(4.000,1.000){2}{\rule{0.964pt}{0.400pt}}
\put(330.0,284.0){\rule[-0.200pt]{2.168pt}{0.400pt}}
\put(356,284.67){\rule{1.927pt}{0.400pt}}
\multiput(356.00,284.17)(4.000,1.000){2}{\rule{0.964pt}{0.400pt}}
\put(347.0,285.0){\rule[-0.200pt]{2.168pt}{0.400pt}}
\put(381,285.67){\rule{2.168pt}{0.400pt}}
\multiput(381.00,285.17)(4.500,1.000){2}{\rule{1.084pt}{0.400pt}}
\put(364.0,286.0){\rule[-0.200pt]{4.095pt}{0.400pt}}
\put(398,286.67){\rule{2.168pt}{0.400pt}}
\multiput(398.00,286.17)(4.500,1.000){2}{\rule{1.084pt}{0.400pt}}
\put(390.0,287.0){\rule[-0.200pt]{1.927pt}{0.400pt}}
\put(424,287.67){\rule{1.927pt}{0.400pt}}
\multiput(424.00,287.17)(4.000,1.000){2}{\rule{0.964pt}{0.400pt}}
\put(407.0,288.0){\rule[-0.200pt]{4.095pt}{0.400pt}}
\put(441,288.67){\rule{1.927pt}{0.400pt}}
\multiput(441.00,288.17)(4.000,1.000){2}{\rule{0.964pt}{0.400pt}}
\put(432.0,289.0){\rule[-0.200pt]{2.168pt}{0.400pt}}
\put(458,289.67){\rule{1.927pt}{0.400pt}}
\multiput(458.00,289.17)(4.000,1.000){2}{\rule{0.964pt}{0.400pt}}
\put(449.0,290.0){\rule[-0.200pt]{2.168pt}{0.400pt}}
\put(483,290.67){\rule{2.168pt}{0.400pt}}
\multiput(483.00,290.17)(4.500,1.000){2}{\rule{1.084pt}{0.400pt}}
\put(466.0,291.0){\rule[-0.200pt]{4.095pt}{0.400pt}}
\put(500,291.67){\rule{2.168pt}{0.400pt}}
\multiput(500.00,291.17)(4.500,1.000){2}{\rule{1.084pt}{0.400pt}}
\put(492.0,292.0){\rule[-0.200pt]{1.927pt}{0.400pt}}
\put(517,292.67){\rule{2.168pt}{0.400pt}}
\multiput(517.00,292.17)(4.500,1.000){2}{\rule{1.084pt}{0.400pt}}
\put(509.0,293.0){\rule[-0.200pt]{1.927pt}{0.400pt}}
\put(543,293.67){\rule{1.927pt}{0.400pt}}
\multiput(543.00,293.17)(4.000,1.000){2}{\rule{0.964pt}{0.400pt}}
\put(526.0,294.0){\rule[-0.200pt]{4.095pt}{0.400pt}}
\put(560,294.67){\rule{1.927pt}{0.400pt}}
\multiput(560.00,294.17)(4.000,1.000){2}{\rule{0.964pt}{0.400pt}}
\put(551.0,295.0){\rule[-0.200pt]{2.168pt}{0.400pt}}
\put(585,295.67){\rule{2.168pt}{0.400pt}}
\multiput(585.00,295.17)(4.500,1.000){2}{\rule{1.084pt}{0.400pt}}
\put(568.0,296.0){\rule[-0.200pt]{4.095pt}{0.400pt}}
\put(602,296.67){\rule{2.168pt}{0.400pt}}
\multiput(602.00,296.17)(4.500,1.000){2}{\rule{1.084pt}{0.400pt}}
\put(594.0,297.0){\rule[-0.200pt]{1.927pt}{0.400pt}}
\put(619,297.67){\rule{2.168pt}{0.400pt}}
\multiput(619.00,297.17)(4.500,1.000){2}{\rule{1.084pt}{0.400pt}}
\put(611.0,298.0){\rule[-0.200pt]{1.927pt}{0.400pt}}
\put(645,298.67){\rule{1.927pt}{0.400pt}}
\multiput(645.00,298.17)(4.000,1.000){2}{\rule{0.964pt}{0.400pt}}
\put(628.0,299.0){\rule[-0.200pt]{4.095pt}{0.400pt}}
\put(662,299.67){\rule{1.927pt}{0.400pt}}
\multiput(662.00,299.17)(4.000,1.000){2}{\rule{0.964pt}{0.400pt}}
\put(653.0,300.0){\rule[-0.200pt]{2.168pt}{0.400pt}}
\put(687,300.67){\rule{2.168pt}{0.400pt}}
\multiput(687.00,300.17)(4.500,1.000){2}{\rule{1.084pt}{0.400pt}}
\put(670.0,301.0){\rule[-0.200pt]{4.095pt}{0.400pt}}
\put(704,301.67){\rule{2.168pt}{0.400pt}}
\multiput(704.00,301.17)(4.500,1.000){2}{\rule{1.084pt}{0.400pt}}
\put(696.0,302.0){\rule[-0.200pt]{1.927pt}{0.400pt}}
\put(721,302.67){\rule{2.168pt}{0.400pt}}
\multiput(721.00,302.17)(4.500,1.000){2}{\rule{1.084pt}{0.400pt}}
\put(713.0,303.0){\rule[-0.200pt]{1.927pt}{0.400pt}}
\put(747,303.67){\rule{1.927pt}{0.400pt}}
\multiput(747.00,303.17)(4.000,1.000){2}{\rule{0.964pt}{0.400pt}}
\put(730.0,304.0){\rule[-0.200pt]{4.095pt}{0.400pt}}
\put(764,304.67){\rule{1.927pt}{0.400pt}}
\multiput(764.00,304.17)(4.000,1.000){2}{\rule{0.964pt}{0.400pt}}
\put(755.0,305.0){\rule[-0.200pt]{2.168pt}{0.400pt}}
\put(789,305.67){\rule{2.168pt}{0.400pt}}
\multiput(789.00,305.17)(4.500,1.000){2}{\rule{1.084pt}{0.400pt}}
\put(772.0,306.0){\rule[-0.200pt]{4.095pt}{0.400pt}}
\put(806,306.67){\rule{2.168pt}{0.400pt}}
\multiput(806.00,306.17)(4.500,1.000){2}{\rule{1.084pt}{0.400pt}}
\put(798.0,307.0){\rule[-0.200pt]{1.927pt}{0.400pt}}
\put(823,307.67){\rule{2.168pt}{0.400pt}}
\multiput(823.00,307.17)(4.500,1.000){2}{\rule{1.084pt}{0.400pt}}
\put(815.0,308.0){\rule[-0.200pt]{1.927pt}{0.400pt}}
\put(849,308.67){\rule{1.927pt}{0.400pt}}
\multiput(849.00,308.17)(4.000,1.000){2}{\rule{0.964pt}{0.400pt}}
\put(832.0,309.0){\rule[-0.200pt]{4.095pt}{0.400pt}}
\put(866,309.67){\rule{1.927pt}{0.400pt}}
\multiput(866.00,309.17)(4.000,1.000){2}{\rule{0.964pt}{0.400pt}}
\put(857.0,310.0){\rule[-0.200pt]{2.168pt}{0.400pt}}
\put(891,310.67){\rule{2.168pt}{0.400pt}}
\multiput(891.00,310.17)(4.500,1.000){2}{\rule{1.084pt}{0.400pt}}
\put(874.0,311.0){\rule[-0.200pt]{4.095pt}{0.400pt}}
\put(908,311.67){\rule{2.168pt}{0.400pt}}
\multiput(908.00,311.17)(4.500,1.000){2}{\rule{1.084pt}{0.400pt}}
\put(900.0,312.0){\rule[-0.200pt]{1.927pt}{0.400pt}}
\put(925,312.67){\rule{2.168pt}{0.400pt}}
\multiput(925.00,312.17)(4.500,1.000){2}{\rule{1.084pt}{0.400pt}}
\put(917.0,313.0){\rule[-0.200pt]{1.927pt}{0.400pt}}
\put(951,313.67){\rule{1.927pt}{0.400pt}}
\multiput(951.00,313.17)(4.000,1.000){2}{\rule{0.964pt}{0.400pt}}
\put(934.0,314.0){\rule[-0.200pt]{4.095pt}{0.400pt}}
\put(968,314.67){\rule{1.927pt}{0.400pt}}
\multiput(968.00,314.17)(4.000,1.000){2}{\rule{0.964pt}{0.400pt}}
\put(959.0,315.0){\rule[-0.200pt]{2.168pt}{0.400pt}}
\put(993,315.67){\rule{2.168pt}{0.400pt}}
\multiput(993.00,315.17)(4.500,1.000){2}{\rule{1.084pt}{0.400pt}}
\put(976.0,316.0){\rule[-0.200pt]{4.095pt}{0.400pt}}
\put(1010,316.67){\rule{2.168pt}{0.400pt}}
\multiput(1010.00,316.17)(4.500,1.000){2}{\rule{1.084pt}{0.400pt}}
\put(1002.0,317.0){\rule[-0.200pt]{1.927pt}{0.400pt}}
\put(1027,317.67){\rule{2.168pt}{0.400pt}}
\multiput(1027.00,317.17)(4.500,1.000){2}{\rule{1.084pt}{0.400pt}}
\put(1019.0,318.0){\rule[-0.200pt]{1.927pt}{0.400pt}}
\put(1053,318.67){\rule{1.927pt}{0.400pt}}
\multiput(1053.00,318.17)(4.000,1.000){2}{\rule{0.964pt}{0.400pt}}
\put(1036.0,319.0){\rule[-0.200pt]{4.095pt}{0.400pt}}
\put(368,113){\usebox{\plotpoint}}
\put(368,113){\circle*{12}}
\put(368.0,113.0){\rule[-0.200pt]{0.400pt}{140.686pt}}
\end{picture}
\end{minipage}
This is unsatisfactory because we assume the masses of the elementary
fermions to be very high. In order to get realistic masses one is
not allowed to
use the strong-coupling limit. Furthermore one should  use a more realistic
propagator instead of the free
propagator and in addition one had to take into account the
polarization cloud. But we are not interested in numerical values of
masses or coupling constants respectively, rather the above discussion
should demonstrate the appearance of \underline{{\sl three}} solution
manifolds which is offered by mixed symmetric spin states.
 \section{Summary and outlook}\label{sec5}
In this paper we have calculated
a special class of spin $1/2$ solutions of generalized three-particle
B.~W.~equations. The reason why we have concentrated
ourselves on solutions with mixed symmetry is the appearance of isospin
doublets (see appendix \ref{b}, table 2) and the appearance of
{\sl three} linearly independent solution
manifolds. The calculation of generalized three-particle B.~W.~equations
resulting from the nonlinear spinor equation (\ref{fg}) is only a first
step in the calculation of the three-subfermion bound state dynamics. It
has already been emphasized in section \ref{sec3} that the
non-diagonal part of (\ref{eeg}) mediates the interactions of the
three-particle states. It has to be shown in a forthcoming paper that the
effective interaction between the mixed symmetric three-particle states and
the two-subfermion composites leads to the inclusion of the three
generations of leptons and quarks. This effective interaction has to
be calculated in the framework of the weak mapping procedure which is a
mathematical tool for calculating effective bound state dynamics \cite{16}.
\section*{Acknowledgement}
The author is very much indebted to Prof.~Dr.~H.~Stumpf for
a critical reading of the manuscript and for many valuable
comments. I would like to thank also Dr.~A.~Buck for
discussions about three-particle problems.
%
%
%
 \section*{Appendix}
 \begin{appendix}
\section{The Young-operators of S(3) which correspond to mixed Symmetries}
\label{a}
In this section we consider those irreducible representation of
the group S(3) which correspond to the Young-diagram $\yd{2mm}$.
Because we have two possible Young-tableaus ($\yta{2mm}\, ,\,
\ytb{2mm}$), the representation is two-dimensional and the
corresponding four Young-operators $C_{ik}$ are defined as \cite{kjs}:
\begin{equation}
   C_{ik} := {1\over 3} \sum_{p \in S(3)} D_{ik}
   \left( p^{-1} \right) P \label{p1.1}
\end{equation}
where $D_{ik}$ are irreducible two-dimensional matrix representations of
the permutation group $S(3)$ and $P$ is an operator representation of
the abstract $S(3)$-element $p$. To be definite we choose \cite{laux}
\def\arraystretch{2}
\begin{equation}
\begin{array}{cc}\def\arraystretch{1} D(e
)=\left(\begin{array}{cc}1&0\\0&1\end{array}\right) &
D(p_{12})=\left(\begin{array}{cc}-1&0\\0&1\end{array}\right)\\ D(p_{13})=
\left(\begin{array}{cc}1/2&-1/2\sqrt{3}\\-1/2\sqrt{3}&-1/2\end{array}\right) &
D(p_{23})=\left(\begin{array}{cc}1/2&1/2\sqrt{3}\\1/2\sqrt{3}&
-1/2\end{array}\right) \\
D(p_{13}\cdot p_{12})
=\left(\begin{array}{cc}-1/2&-1/2\sqrt{3}\\1/2\sqrt{3}&-1/2
\end{array}\right)& D(p_{12}\cdot p_{13})=
\left(\begin{array}{cc}-1/2&1/2\sqrt{3}\\-1/2\sqrt{3}&-1/2\end{array}\right)
\end{array}\label{p1.2}
\end{equation}
\def\arraystretch{1}
We have denoted the transpositions which interchange $j,l$
by $p_{jl}$. From (\ref{p1.1}) and (\ref{p1.2}) we obtain
\def\arraystretch{1.5}
\begin{equation}
\begin{array}{lclcl}

   C_{11} &\, =\,& {1\over 3} \left( 2 + P_{13} + P_{23}\right)
           {1\over 2} \left( 1 - P_{12} \right)&\,=\,& {1\over 2}
\left( 1 - P_{12} \right) {1\over 3} \left( 2 + P_{13} +
P_{23}\right)  \\
    C_{22} &\,=\,& {1\over 3} \left( 2 - P_{13} - P_{23}\right)
           {1\over 2} \left( 1 + P_{12} \right)&\,=\,&
            {1\over 2}
\left( 1 + P_{12} \right) {1\over 3} \left( 2 - P_{13} -
P_{23}\right)   \\
    C_{12} &\,=\,& {{\sqrt 3}\over 3} \left( P_{23} - P_{13} \right)
           {1\over 2} \left( 1 + P_{12} \right)&\, =\,& {1\over 2}
\left( 1 - P_{12} \right) {{\sqrt 3}\over 3} \left( P_{23} -
P_{13} \right)  \\
    C_{21} &\, =\,& {{\sqrt 3}\over 3} \left( P_{23} - P_{13} \right)
           {1\over 2} \left( 1 - P_{12} \right)&\, =\,& {1\over 2}
\left( 1 + P_{12} \right){{\sqrt 3}\over 3} \left( P_{23} -
P_{13} \right)
\end{array}\label{p2.1}
\end{equation}
\def\arraystretch{1}
where $P_{ik}$ is an operator representation of the S(3)-element
$p_{ik}$. In \cite{kjs} the following properties of the $C_{ik}$
are proven
\begin{eqnarray}
C^a\, C_{ik}&=& C^s\, C_{ik}=0 \label{I14.2} \\
C^+_{ik}&=&C_{ki} \label{p3.1}\\
C_{ik}\cdot C_{lj}&=&\delta_{kl}\, C_{ij} \label{p3.2}\\
P\, C_{ik}&=&\left( D^T(p)\cdot C\right)_{ik} \label{p2.2}
\end{eqnarray}
The matrix $D^T$ is the transposed of the matrix $D$. In
this paper the following applications of (\ref{p2.2}) are used:
\begin{eqnarray}
P_{23}\, C_{11} &=& \frac{1}{2} C_{11}+\frac{1}{2}\sqrt{3}
C_{21} \label{p3.3}\\
P_{13}\, C_{11} &=& \frac{1}{2} C_{11}-\frac{1}{2}\sqrt{3}
C_{21} \label{p3.4}
\end{eqnarray}
 \section{Isospin States with mixed Symmetry} \label{b}
Due to the global SU(2)$\times$U(1) form invariance of the spinor
theory, we can classify the three-particle states $\left|
a\right>$ according to SU(2) and U(1) quantum numbers, ignoring
the possibility of symmetry breaking.
Therefore, we require the state $\left| a\right>$ to fulfil the
equations
\begin{equation}
\renewcommand{\arraystretch}{1.3}
\begin{array}{rrc}
T^k T^k \left| a\right> = t (t+1) \left| a
\right>,  &  \,\,
T^3 \left| a\right>  =  t^z \left| a\right>, & \, \,
F\left| a\right> = f \left| a\right>,  \\
Q \left| a\right>  =  q \left| a\right>, &   &
\end{array} \label{*a}
\end{equation}
where $T^k$ are the generators of the SU(2) transformations which satisfy
\begin{equation}
\left[ T^k , \psi_I \right] = - T^k_{I I'} \psi_{I'} \quad
{\rm with}\quad T^k_{I I'}:=\frac{1}{2}\left( \begin{array}{lr}
\sigma^k & 0 \\ 0 & (-)^k \sigma^k \end{array} \right)_{\kappa
\kappa '}\delta_{\alpha \alpha'}\, \delta ({\bf r}-{\bf r}')
\quad . \label{a1.2}
\end{equation}
The U(1) transformations which are generated by
$F,Q$ can be directly read off from the
relations
\jot3mm
\begin{eqnarray}
\left[ F , \psi_I \right] = - F_{I I'} \psi_{I'}& \quad
{\rm with} \quad & F_{I I'}:= \frac{1}{3}\,\left( \begin{array}{lr}
\1 & 0 \\ 0 & -\1 \end{array} \right)_{\kappa
\kappa '}\delta_{\alpha \alpha'}\, \delta ({\bf r}-{\bf r}')\, , \\
\left[ Q , \psi_I \right] = - Q_{I I'} \psi_{I'}& \quad
{\rm with} \quad & Q_{I I'}:= \frac{1}{3}\left( \begin{array}{cccc}
2 & 0 & 0& 0\\ 0 &-1&0&0 \\ 0&0&-2&0\\ 0&0&0&1 \end{array} \right)_{\kappa
\kappa '}\delta_{\alpha \alpha'}\, \delta ({\bf r}-{\bf r}')\, , \\
\end{eqnarray}
\jot0mm
The SU(2) quantum numbers $t,t_z$ are called isospin quantum
numbers whereas the U(1) quantum numbers $f,q$ are called
fermion number and charge.
Obviously, the transformations generated by $T^3,F,Q$ are not
independent. As a consequence, the relation $q=t^z+f/2$ holds. \\ \\[-3mm]
 From (\ref{*a}) we get
\jot1mm
\begin{eqnarray}
\lefteqn{
\frac{9}{4}\, \,\varphi_{I_1 I_2 I_3}\, + \, 2\left[
T^k_{I_1 I} T^k_{I_2 I'}\,\,\varphi_{I I'I_3} \right. }\nonumber \\
& &
\left.  + \, T^k_{I_1 I}
T^k_{I_3 I'}\,\,\varphi_{I I_2 I'}\, +\, T^k_{I_2 I} T^k_{I_3
I'}\,\,\varphi_{I_1 I I'} \right] = t(t+1)\,\, \varphi_{I_1 I_2
I_3}\quad , \label{j1.3}
\end{eqnarray}
\jot0mm
and:
\begin{eqnarray}
T^3_{I_1 I}\,\,\varphi_{I I_2 I_3}\, + \, T^3_{I_2
I}\,\,\varphi_{I_1 I I_3} \, + \, T^3_{I_3 I}\,\,\varphi_{I_1 I_2 I}
& = & t^z \, \varphi_{I_1 I_2 I_3}  \label{j4.0} \\
F_{I_1 I}\,\,\varphi_{I I_2 I_3}\, + \, F_{I_2
I}\,\,\varphi_{I_1 I I_3} \, + \, F_{I_3 I}\,\,\varphi_{I_1 I_2 I}
& = & f \, \varphi_{I_1 I_2 I_3}  \label{j4.1} \\
Q_{I_1 I}\,\,\varphi_{I I_2 I_3}\, + \, Q_{I_2
I}\,\,\varphi_{I_1 I I_3} \, + \, Q_{I_3 I}\,\,\varphi_{I_1 I_2 I}
& = & q \, \varphi_{I_1 I_2 I_3}  \label{j5.1}
\end{eqnarray}
If we substitute the ansatz (\ref{I8.2}) into
(\ref{j1.3})-(\ref{j5.1}), we get equations for the
isospin-superspin part $C_{11}\left| \Theta^j\right>\, ,\,
C_{22}\left|\Theta^j\right>$
i.~e.~ we have to discuss isospin states with mixed symmetry,
characterized by the Young-diagram $\yd{2mm}$. We restrict ourselves
to those isospin states which correspond to the Young-tableau
$\ytb{2mm}$. These states are eigenstates to $C_{11}$ or
$C_{22}$ resp.~with eigenvalues 1 or 0 resp.. The states which
correspond to the tableau $\yta{2mm}$ can be generated from the
$\ytb{2mm}$-states with the help of the operator $C_{21}$. The
number of independent tensors in $n$ dimensions which correspond
to a special tableau can be calculated according to
\cite{littlewood}. For mixed symmetry we have
\begin{equation}
\frac{\ytn{7mm}}{\ytc{7mm}}=\frac{(n+1)n(n-1)}{3}
\end{equation}
In our case ($n=4$) we have 20 linear independent isospin states
corresponding to $\ytb{2mm}$. For the multiplicity of the states
we have
\begin{displaymath}
20=\underbrace{2}_{\rm states\, +\, charge\,\, conjugated\,\, states}\cdot
\left( 1\cdot\underbrace{4}_{\rm
quartet}+3\cdot\underbrace{2}_{\rm doublet}\right)
\end{displaymath}
However, we must take into account the condition \\
\parbox{8cm}{\begin{displaymath}
(\gamma_5 )_{\kappa_2\kappa_3}\,
(C_{11}\Theta^j)_{\kappa_1\kappa_2\kappa_3} =0
\end{displaymath}}\hspace*{\fill}(\ref{I11.1})\\
(see section \ref{sec4}). Because (\ref{I11.1}) has
an open index $\kappa$ we have four constraints which
restrict the number of states to $20-4=16$.
These $16$ states can be classified according to SU(2) and U(1)
quantum numbers i.~e.~they can be chosen to fulfill
(\ref{j1.3})-(\ref{j5.1}). In order to present these states in a
readable form, we introduce the following definitions
\begin{equation}
T^+_{\kappa}:=\delta_{1\kappa}\,\, ,\,\,
T^-_{\kappa}:=\delta_{2\kappa} \, \, ,\,\,
V^-_{\kappa}:=\delta_{3\kappa}\,\, ,\,\,
V^+_{\kappa}:=\delta_{4\kappa}
\end{equation}
The U(1), SU(2) quantum numbers of the subfermions are
given in the following table: \\[0.4cm]
\parbox{5.7cm}{
\begin{tabular}{|l|c|c|c|c|}
\hline
  &$T^+$ &$T^-$& $V^+$& $V^-$ \\
\hline
$t$ &  1/2    & 1/2  & 1/2  & 1/2  \\
\hline
$t_z$ &  1/2    &-1/2  &1/2&-1/2   \\
\hline
$f$&1/3  &1/3 &-1/3&-1/3 \\
\hline
$q$ &2/3 &-1/3&1/3 &-2/3  \\
\hline
\end{tabular}} Table 1
\\[0.23cm]
For simplicity we we omit the index $\kappa$. For instance the
expression $T^+V^-T^-$ is an abbreviation for
$T^+_{\kappa_1}V^-_{\kappa_2}T^-_{\kappa_3}=
\delta_{1\kappa_1}\delta_{3\kappa_2}\delta_{2\kappa_3}$. With
these definitions, the 16 isospin states which fulfill
(\ref{I11.1}) are given by
\jot1.5mm
\begin{eqnarray*}
\Theta^1&:=&\frac{1}{\sqrt{2}}\left( T^+T^-T^+ -T^-T^+T^+\right)\\
\Theta^2&:=&\frac{1}{\sqrt{2}}\left( T^+T^-T^- -T^-T^+T^-\right)\\
\Theta^3&:=&\frac{1}{\sqrt{2}}\left( V^+V^-V^+ -V^-V^+V^+\right)\\
\Theta^4&:=&\frac{1}{\sqrt{2}}\left( V^+V^-V^- -V^-V^+V^-\right)\\
\Theta^5 &:=&\frac{1}{\sqrt{6}}\Big(
T^+T^-V^+-T^-T^+V^++V^+T^-T^+-T^-V^+T^+ \\
&&+V^-T^+T^+ -T^+V^-T^+\Big) \\
\Theta^6 &:=&\frac{1}{\sqrt{6}}\Big(
T^-T^+V^--T^+T^-V^-+V^-T^+T^--T^+V^-T^- \\
&&+V^+T^-T^- -T^-V^+T^-\Big) \\
\Theta^7 &:=&\frac{1}{\sqrt{6}}\Big(
V^+V^-T^+-V^-V^+T^++T^+V^-V^+-V^-T^+V^+ \\
&&+T^-V^+V^+ -V^+T^-V^+\Big) \\
\Theta^8 &:=&\frac{1}{\sqrt{6}}\Big(
V^-V^+T^--V^+V^-T^-+T^-V^+V^--V^+T^-V^- \\
&&+T^+V^-V^- -V^-T^+V^-\Big) \\
\Theta^9&:=&\frac{1}{\sqrt{2}}\Big( T^+V^+T^+-V^+T^+T^+\Big) \\
\Theta^{10}&:=&\frac{1}{\sqrt{6}}\Big(
T^+V^+T^--V^+T^+T^--T^+V^-T^++V^-T^+T^+\\
&&+T^-V^+T^+-V^+T^-T^+ \Big) \\
\Theta^{11}&:=&\frac{1}{\sqrt{6}}\Big(
T^-V^-T^+-V^-T^-T^+-T^-V^+T^-+V^+T^-T^-\\
&&+T^+V^-T^--V^-T^+T^- \Big) \\
\Theta^{12}&:=&\frac{1}{\sqrt{2}}\Big( T^-V^-T^--V^-T^-T^-\Big) \\
\Theta^{13}&:=&\frac{1}{\sqrt{2}}\Big( V^+T^+V^+-T^+V^+V^+\Big) \\
\Theta^{14}&:=&\frac{1}{\sqrt{6}}\Big(
V^+T^+V^--T^+V^+V^--V^+T^-V^++T^-V^+V^+\\
&&+V^-T^+V^+-T^+V^-V^+ \Big) \\
\Theta^{15}&:=&\frac{1}{\sqrt{6}}\Big(
V^-T^-V^+-T^-V^-V^+-V^-T^+V^-+T^+V^-V^-\\
&&+V^+T^-V^--T^-V^+V^- \Big) \\
\Theta^{16}&:=& \frac{1}{\sqrt{2}}\Big( V^-T^-V^--T^-V^-V^-\Big)
\end{eqnarray*}
\jot0mm
The isospin states $\left| j\right>$ which correspond to the matrix elements
\newline
$\Theta^j\equiv
\Theta^j_{\kappa_1\kappa_2\kappa_3}=\left<
\kappa_1\kappa_2\kappa_3\right| \left. j\right>$
fulfill the relation $C_{11}\left| j\right> =\left| j\right> $
and are normalized according to
\begin{equation}
\left< j'\right| C_{11}\left| j\right>=
\left< j'\right| \left. j\right> =\delta_{jj'}\label{norm}\quad .
\end{equation}
The verification of (\ref{I11.1}) is best done with the representation
\begin{displaymath}
\gamma_5=T^+V^-+V^-T^++V^+T^-+T^-V^+
\end{displaymath}
The quantum numbers of the above isospin states are
summarized in the following table \\ \\
\parbox{12.8cm}{
\renewcommand{\arraystretch}{1.2}
\begin{tabular}{|l|c|c|c|c||c|c|c|c|}
\hline
$\left| j\right>$  & $\left| 1\right>$ &$\left|
2\right>$&$\left| 3\right>$& $\left| 4\right>$ & $\left|
5\right>$ & $\left| 6\right>$ & $\left| 7\right>$  &$\left|
8\right>$ \\
\hline
$t$ &  1/2    & 1/2  & 1/2  & 1/2  &  1/2  &  1/2  &  1/2 &
 1/2  \\
\hline
$t^z$ &  1/2    &-1/2  &1/2&-1/2   &1/2  &-1/2&1/2  &-1/2  \\
\hline
$f$&1  & 1   &-1&-1&1/3  & 1/3 &-1/3 &-1/3  \\
\hline
$q$& 1    & 0 &0&-1&2/3&-1/3 &1/3 &-2/3 \\
\hline
\multicolumn{9}{|c|}{Interpretation:} \\
\hline
$\omega_3$& $e^+$    & $\bar{\nu_e}$ &$\nu_e$&$e^-$&$u$&$d$ &$\bar{d}$
&$\bar{u}$ \\
\hline
$\omega_1$& $\mu^+$ & $\bar{\nu_\mu}$ &$\nu_\mu$&$\mu^-$&$c$&$s$ &$\bar{s}$
&$\bar{c}$ \\
\hline
$\omega_2$& $\tau^+$ & $\bar{\nu_\tau}$ &$\nu_\tau$&$\tau^-$&$t$&$b$ &$\bar{b}$
&$\bar{t}$ \\
\hline
\end{tabular}\\[0.3cm]
\begin{tabular}{|l|c|c|c|c||c|c|c|c|}
\hline
$\left| j\right>$  & $\left| 9\right>$ &$\left|
10\right>$&$\left| 11\right>$& $\left| 12\right>$ & $\left|
13\right>$ & $\left| 14\right>$ & $\left| 15\right>$  &$\left|
16\right>$ \\
\hline
$t$ &  3/2    & 3/2  & 3/2  & 3/2  &  3/2  &  3/2  &  3/2 &
 3/2  \\
\hline
$t^z$ &  3/2    &1/2  &-1/2&-3/2   &3/2  &1/2&-1/2  &-3/2  \\
\hline
$f$&1/3  & 1/3   &1/3&1/3&-1/3  & -1/3 &-1/3 &-1/3  \\
\hline
$q$& 5/3    & 2/3 &-1/3&-4/3&4/3&1/3 &-2/3 &-5/3 \\
\hline
\end{tabular}}\hfill Table 2\\[0.42cm]
%
%
At this stage of calculation the name isospin, fermion number
and charge do not imply any physical interpretation
of the states. Rather these quantum numbers reflect the symmetry
of eqn.~(\ref{d1.1}) and serve as bookkeeping indices only. In order
to determine the phenomenological quantum numbers of the three-subfermion
bound states, the interaction with other bound state
particles must be taken into account. However, anticipating the results
of a forthcoming paper in which the effective interaction of these
three-particle states with the two-fermion composites will be calculated,
we may identify the quantum numbers $t,t^z,f,q$ as well as the hypercharge
$y=t^z+\frac{f}{2}$ with the phenomenological quantum numbers
of leptons and quarks.%
 \end{appendix}
%
%
%

%
%
\end{document}